\documentclass[11pt,a4paper]{article}
\usepackage{upgreek}
\usepackage[english]{babel}	
\usepackage[utf8]{inputenc}	
\usepackage{lmodern}	

\usepackage[a4paper,lmargin={2.5cm},rmargin={2.5cm},tmargin={2.5cm},bmargin = {2.5cm}]{geometry}	
\usepackage{pdfpages}

\usepackage{amsmath}	
\usepackage{amssymb}	
\usepackage{dsfont}	
\usepackage{mathrsfs}

\usepackage{paralist}
\usepackage{cite}

\usepackage{graphicx}
\usepackage[export]{adjustbox}

\graphicspath{{Bilder/}}	
\usepackage{enumerate}
\usepackage{caption}
\usepackage{subcaption}	
\usepackage{float,rotating}
\usepackage{array,booktabs}

\usepackage{multirow}
\usepackage{hhline}

\usepackage{verbatim}

\usepackage{bbold}

\usepackage[separate-uncertainty = false,multi-part-units=single]{siunitx}

\usepackage[version=3]{mhchem}

\begin{document}
	\begin{titlepage}
		\thispagestyle{empty} %

		\begin{center}
			\vspace{3.0cm}
			
			\textbf{ 
			\LARGE Non-parametric model-based estimation of the effective reproduction number for SARS-CoV-2} \\ \vspace{1.5cm}
			
			\textsc{\Large Hermes Jacques$^{a,b,*}$, Rosenblatt Marcus$^{a,b}$,\\ Tönsing Christian$^{a,b,c}$, Timmer Jens$^{a,b,c}$} \\[2ex]
			$^a$Institute of Physics, University of Freiburg, Hermann-Herder-Str. 3, 79104 Freiburg, Germany\\
$^b$Freiburg Center for Data Analysis and Modelling (FDM), University of Freiburg, Ernst-Zermelo-Str. 1, 79104 Freiburg, Germany\\
$^c$Centre for Integrative Biological Signalling Studies (CIBSS), University of Freiburg, Schänzlestr. 18, 79104 Freiburg, Germany\\
			$^*$ Corresponding author: jacques.hermes@fdm.uni-freiburg.de\\
			\today
		\end{center}
	
\vfill

\begin{abstract}
Viral outbreaks, such as the current COVID-19 pandemic, are commonly described by compartmental models by means of ordinary differential equation (ODE) systems. The parameter values of these ODE models are typically unknown and need to be estimated based on accessible data. In order to describe realistic pandemic scenarios with strongly varying situations, these model parameters need to be assumed as time-dependent. While parameter estimation for the typical case of time-constant parameters does not pose larger issues, the determination of time-dependent parameters, e.g.~the transition rates of compartmental models, remains notoriously difficult, in particular since the function class of these time-dependent parameters is unknown.

In this work, we present a novel method which utilizes the Augmented Kalman Smoother in combination with an Expectation-Maximization algorithm to simultaneously estimate all time-dependent parameters in an SIRD compartmental model. This approach only requires incidence data, but no prior knowledge on model parameters or any further assumptions on the function class of the time-dependencies. In contrast to other approaches for the estimation of the time-dependent reproduction number, no assumptions on the parameterization of the serial interval distribution are required. With this method, we are able to adequately describe COVID-19 data in Germany and to give non-parametric model-based time course estimates for the effective reproduction number. 

\end{abstract}
		

	\end{titlepage}

\section{Introduction}

Over the past 50 years, the rate of emerging infectious diseases, especially zoonoses and vector-borne diseases, has increased drastically \cite{jones2008global, world2014vector}. A better understanding of the dynamics of such diseases demands for a quantitative description by means of mathematical models \cite{ hethcote2000mathematics, murray2002mathematical, brauer2012mathematical, tonsing2018profile, abboubakar2021mathematical}. The novel SARS-CoV-2 caused a global pandemic in 2020/21 and gave rise to a multitude of modeling efforts such as early situation assessments \cite{barbarossa2020modeling, dehning2020inferring, bai2020rapid}, analysis of the effectiveness of non-pharmaceutical interventions \cite{ferguson2020report, jarvis2020quantifying, latsuzbaia2020evolving}, the impact of undetected cases \cite{ivorra2020mathematical, contento2021integrative} or agent-based model descriptions \cite{rockett2020revealing, truszkowska2021high, hoertel2020stochastic}. 

A common approach to capture the dynamics of a viral disease spread is to utilize the classical Susceptible-Infected-Recovered (SIR) compartmental model and to estimate its parameters from recorded data \cite{capaldi2012parameter, kermack1991contributions, chen2021numerical}. These model parameters are typically unknown and differ for every disease, depending on the pathogen and its contagiousness \cite{bani2012reproduction, tonsing2018profile}. For many infectious diseases, extensions of the classical SIR structure have been proposed in order to yield a more appropriate description of the infection dynamics. Such adaptations commonly distinguish between deceased (D) and recovered (R) individuals or add an intermediate exposed compartment (E) in which individuals are infected but not yet infectious. Introducing even more compartments and interactions in such models might add detail to the description of the infection dynamics, however, parameter estimation in overly complex models is hampered, because of the limited amount of available data. Furthermore, such models typically suffer from parameter-identifiability issues due to over-parameterizations \cite{wieland2021structural}.

Many SIR-like model structures have been proposed and investigated for the COVID-19 disease \cite{barbarossa2020modeling, massonis2020structural, godio2020seir, raimundez2021covid}. 
Comprehensive model structures aim to cover all relevant features of the SARS-CoV-2 pandemic, such as for example asymptomatic infections, undetected infections or different hospitalization states \cite{khailaie2021development}. Despite the general interest in each of the individual parameters in such detailed models, an essential measure for the  transmission potential of a disease in all epidemic models is the effective reproduction number $\mathcal R_t$, which is given by the average number of secondary infections caused by an infectious individual. 
It may vary over time, yields information about the current status of an outbreak, and can for example indicate the turning point of an exponential growth of infections. Its estimation from data can be achieved via epidemic models with time-dependent parameters and has been a key task for scientists to inform e.g.\ governmental decision makers during the SARS-CoV-2 pandemic.

In many countries, time series of the SARS-CoV-2 infection numbers showed multiple epidemic waves \cite{rahmandad2021behavioral, prodanov2021analytical}, similar to temporal changes in the infection numbers as e.g.\ for seasonal influenza \cite{yaari2013modelling}.
To capture these dynamics of the recorded infection case numbers in an SIR-like model, the model parameters need to be time-dependent \cite{jo2020analysis, hong2020estimation, kolokolnikov2021law}.
These time-dependencies of e.g.\ infection rate parameters not only cover seasonal effects \cite{capistran2009parameter, merow2020seasonality}, but also behavioral changes in the population and the impact of other non-pharmaceutical interventions.
The latter might be modeled straightforwardly by a specific step-function which lowers the infection-rate parameter e.g.\ on the date when governmental interventions come into force \cite{dehning2020inferring}. 
Such models enable the comparison of non-pharmaceutical intervention strategies, but highly depend on the chosen class of time-dependent functions and are less suitable to cover behavioral adaptions in the population.
Choosing the appropriate function classes for these time-dependent model parameters and inferring them from recorded data is closely related to what is generally called input estimation for ordinary differential equation (ODE) models, a task that still needs to be solved in particular for non-linear high-dimensional systems \cite{kaschek2012variational, engelhardt2016learning, villaverde2019full, kreutz2020new}.
In contrast, in the context of infectious disease models, a recently published approach does not require such a particular function class definition, but instead is able to estimate the time dependency for one single model parameter by means of a Kalman Smoother method \cite{arroyo2021tracking}. 

The here presented work extends the function class-free approach for time-dependencies of all dynamical parameters in an SIRD model. 
The method consists of an Augmented Kalman Smoother (AKS) in combination with an Expectation-Maximization (EM) algorithm for the estimation of multiple time-dependent model parameters solely from incidence data. 
In particular, our approach does not require any additional prior knowledge about the remaining model parameters or total and initial numbers in the individual population compartments. We apply the algorithm to the SARS-CoV-2 incidence data from winter 2020 to summer 2021 in Germany and are able to estimate the time-dependencies of all model parameters as well as reporting online and real-time estimates of the time-dependent effective reproduction number~$\mathcal R_t$.

\section{Methods}\label{methods}

\subsection{SIRD model and reparameterization}\label{SIRD model}

The spread of an infectious disease within a population is frequently described by the SIR model \cite{kermack1991contributions} consisting of the three compartments: Susceptible $S$, infected $I$ and removed $R$. The third compartment is split up into recovered $R$ and deceased $D$ resulting in the SIRD model \cite{keeling2011modeling} whose time-continuous evolution is given by the following system of ODEs:
\begin{align}
\begin{split}
\dot{S} &= -\beta\cdot\frac{S\cdot I}{N_0}  \label{SIRD} \\
\dot{I} &= \beta\cdot\frac{S\cdot I}{N_0} -\gamma \cdot I-\theta \cdot I \\
\dot{R} &= \gamma \cdot I \\
\dot{D} &= \theta \cdot I \,.
\end{split}
\end{align}
The SIRD model comprises three time-constant transition rate parameters $\beta$, $\gamma$ and $\theta$ which correspond to the infection rate, recovery rate and mortality rate, respectively. The initial values of the four compartments at $t=0$ are denoted by $S(0)=S_0$, $I(0)=I_0$, while $R(0)=0$ and $D(0)=0$ so that the total number of individuals is $N_0 = S_0 + I_0$.

At the beginning of an epidemic, i.e.\ when $S_0\approx N_0$, the \textit{basic reproduction number} $\mathcal{R}_0$ of the disease, corresponding to the average number of secondary cases, can be calculated directly from the model parameters as $\mathcal{R}_0 = \beta/(\gamma + \theta)$.  
During the course of an epidemic, seasonal effects for the viral transmission and behavioral changes in the population, for example increased hygiene standards or other non-pharmaceutical interventions, are reflected by time-dependencies in the infection rate~$\beta$.
Such effects may also result in epidemic scenarios with multiple waves of infections, as observed e.g. for seasonal influenza or COVID-19.
In order to account for temporal changes in the model parameters, the transition rates $\beta_t$, $\gamma_t$ and $\theta_t$ are assumed to vary over time.  
This further allows for the definition of a time-dependent \textit{effective reproduction number} \cite{nishiura2009effective}
\begin{align}
\mathcal{R}_t&= \frac{\beta_t}{\gamma_t+\theta_t} \cdot \frac{S(t)}{N_0} \label{R}
\end{align}
which depends on the change of susceptibles $S(t)$ over time and on the time courses of the three parameters $\beta_t$, $\gamma_t$ and $\theta_t$.
Since for $\mathcal{R}_t>1$ or $\mathcal{R}_t<1$ the number of infected persons is increasing, or decreasing, respectively, the effective reproduction number is a characteristic measure for the current status of an epidemic. 

To simplify the system for the estimation of $\mathcal{R}_t$, we substitute Equation~\eqref{R} into Equation~\eqref{SIRD}, yielding the reparameterized form of the SIRD model:
\begin{align}
\begin{split}
\dot{S} &= - \mathcal{R}_t \cdot (\gamma_t+\theta_t) \cdot I  \label{SIRD_repar}\\
\dot{I} &= (\mathcal{R}_t-1) \cdot (\gamma_t+\theta_t) \cdot I \\
\dot{R} &= \gamma_t \cdot I \\
\dot{D} &= \theta_t \cdot I.
\end{split}
\end{align} 
Note that by means of this transformation, the compartment of susceptibles $S$ is no longer appearing on the right-hand-side of Equation~\eqref{SIRD_repar}, therefor the differential equation for $S$ could be neglected without affecting the solution of the remaining equations.

Parameter estimation in non-linear ODE models with time-constant parameters is typically tackled by efficient numerical optimization algorithms that minimize the discrepancy between recorded data and model trajectories by tuning the parameter values, e.g.~by maximum-likelihood estimation \cite{raue2013lessons, raue2015data2dynamics, stapor2018pesto, loos2018hierarchical, kaschek2019dynamic}.
In the presented setting of changing transmission rates during an infectious disease outbreak, also the time-dependency of the model parameters itself is unknown and needs to be estimated from data.
This task is known as input estimation for ODE models in general.
However, most methods are limited in their restriction to a certain function class for these time-dependent model parameters, e.g.~step-functions, splines or combinations of sustained and transient response functions \cite{kreutz2020new,schelker2012comprehensive}.
In contrast, we present a function class-free approach based on Kalman Filter and Smoother methods that is able to estimate the dynamics of unobserved model states and after some extensions also the time-dependencies of the parameters.

\subsection{Estimation of time-dependent parameters in non-linear ODE models via the Augmented Kalman Smoother}
The Kalman Filter is an iterative algorithm that is used to estimate time courses of unobserved states of a linear model from noisy observations over time. 
The corresponding Kalman Smoother is a two-step algorithm that uses the Kalman Filter as forward-pass followed by a backwards-pass procedure of the same kind. 
The step-wise linearizations of the Kalman Filter and Smoother for non-linear models is covered by the so-called Extended Kalman Filter and Smoother.
Ultimately, the \textit{Augmented Kalman Smoother} (AKS) combines the latter with the ability to consider time-dependent parameters.

\subsubsection*{State-space model}
The Kalman Filter and its extensions operate in the time-discrete state-space representation of a given model. It has the general form
\begin{align}
\begin{split}
	x_{k+1} &= b(x_k)+ \epsilon_k   \\ \label{state_space}
	y_k &= H\cdot x_k + \eta_k   \,,
\end{split}
\end{align}
for discrete time points $k$ of the $n$-dimensional process state vector $x_k$ and the $d$-dimensional vector of observables $y_k$, with $n > d$. The transition of the state vector $x$ from time point $k$ to $k+1$ is given by the transition function $b$ with assumed additive Gaussian normally distributed process noise $\epsilon_k\sim \mathcal{N}(0,Q)$ with covariance matrix $Q$. 
Since not all states are observed directly, the observation matrix $H$ defines the mapping between the state vector $x$ and the observable vector $y$.
Likewise to the process states, additive Gaussian distributed measurement noise $\eta_k \sim \mathcal{N}(0,R)$ with covariance matrix $R$ is assumed for the observations $y_k$.

The model structure of a given ODE model~$\dot x=f(x,p)$ with $m$ parameters~$p$ can be transferred into its time-discrete state-space formulation, where the solutions of the ODE correspond to the states~$x$ of the state-space and the right-hand side~$f$ of the ODE is translated into time-discrete update rules for the transition function~$b$. 
Furthermore, the observation matrix~$H$ needs to be specified in order to define the link from the data in vector~$y$ to the respective states from vector~$x$.

\subsubsection*{Augmented Kalman Smoother}

For the estimation of unobserved process states in Equation~\eqref{state_space} in non-linear models, the Extended Kalman Smoother is used \cite{kalman1960new}. 
It assumes time-constant parameters and requires all model parameter values as input.
In order to estimate time-dependent parameters, we utilize an extension of the Extended Kalman Filter and Smoother algorithms and call it Augmented Kalman Smoother (AKS) in correspondence to the previously reported Augmented Kalman Filter \cite{carrassi2011state, sun2008extended}. 
It essentially follows the formulation of the Extended Kalman Smoother (EKS) with the distinction of introducing additional process states for each time-dependent model parameter. 
It therefore only requires appropriate initial values of the states and parameters as input that are subject to change over time.
As a consequence, the AKS not only outputs the estimates for the time courses of the unobserved process states, but also yields estimates for the time courses of the time-dependent model parameters. Moreover, the AKS method estimates the covariance matrices $P$ of the state-space vector $x$ of the analyzed process in the same iteration and is initialized at time point $k=1$. The process state vector $x_0$ and the process state covariance matrix $P_0$ at time point $k=0$ need to be provided. 

The first step of the algorithm is the so-called \textit{prediction step}
\begin{align}
x_{k\vert k-1} &= b(x_{k-1\vert k-1}) \label{pred}\\
P_{k\vert k-1} &= F_{k} \cdot P_{k-1\vert k-1} \cdot F_{k}^\text{T}+Q.
\end{align}
Here $x_{k\vert k-1}$ and $P_{k\vert k-1}$ correspond to the predicted state vector and covariance matrix at time $k$ based on the $k-1$ previous data points.
Thus, the first predicted state vector $x_{1|0}$ is calculated from the transition function $b$ applied to the state vector $x_0$. 
As part of the step-wise linearization, the Jacobian of $b$ at point $x_{k-1 \vert k-1}$ is calculated as
\begin{align}
	F_k &= \frac{\partial b}{\partial x}\biggr\rvert_{x_{k-1\vert k-1}}\,.
\end{align} 
The second step of the algorithm is the \textit{update step}, where the predicted state vector and covariance matrix are updated using the noisy measurement $y_k$:
\begin{align}
K_k &= P_{k\vert k-1} \cdot H_k^\text{T} \cdot \left(H_k \cdot P_{k\vert k-1} \cdot H_k^\text{T} + R\right)^{-1} \label{gain}\\
x_{k\vert k}& = x_{k\vert k-1} + K_k \cdot \left({y}_k - H_k \cdot x_{k\vert k-1}\right)\label{up}\\
P_{k\vert k} &= P_{k\vert k-1} - K_k \cdot H_k\cdot P_{k\vert k-1}\label{up1}
\end{align}
The Kalman gain $K_k$ is calculated from the the measurement covariance matrix $R$ and the predicted covariance matrix $P_{k\vert k-1}$ of the process. Since it can be interpreted as the relative weight between the model predictions and the noisy measurements, a large Kalman gain implies that more weight for the next prediction is placed on the measurements. 
In such a case, the uncertainties in the model predictions are considered to be larger than the measurement uncertainties and vice-versa.
The predictions are in turn updated according to Equation\,\eqref{up} and \eqref{up1}.
To close the cycle of the algorithm, the updated state vector ${x}_{k\vert k}$ and covariance matrix $P_{k\vert k}$ are used as input for the following prediction step. This cycle is repeated until the end of the observed time series is reached.

Equations\,\eqref{pred}-\eqref{up1} define the filter, which performs a forward estimation of state-space vector $x$ and its covariance matrix $P$. The shortcoming of the filtering method is that the estimate ${x}_{k\vert k}$ is only based on the $k-1$ previous data points. In consequence, the filter's estimates at the beginning of the time series are based on less information than those at the end. Furthermore, the estimates always lag one step behind the data. To circumvent this issue, the filter is typically combined with a smoother, which is commonly also called Rauch–Tung–Striebel Smoother \cite{rauch1965maximum}.
While the filter runs along the data from $k=1$ to $k=N$, the smoother runs backwards from $k=N-1$ to $k=0$. It is initialized with the last updated estimates of the filter ${x}_{N\vert N}$ and $P_{N\vert N}$. It does not directly use the noisy measurements $y_k$ but rather compares the estimates and predictions that result from the filter to generate smoothed estimates at each time point $k$ which are based on all available data:
\begin{align}
B_k &= P_{k\vert k} \cdot F_k^\text{T} \cdot P_{k+1\vert k}^{-1}\\
x_{k\vert N} &= x_{k\vert k} + B_k \cdot \left(x_{k+1\vert N}- x_{k+1\vert k}\right)\\
P_{k\vert N } &= P_{k\vert k} + B_k \cdot \left(P_{k+1\vert N}- P_{k+1\vert k}\right) \cdot B_k^\text{T}\label{cov}
\end{align}
Here, $B_k$ is the smoother gain which has a similar function to the Kalman gain $K_k$ in Equation\,\eqref{gain}. As a final result, the AKS algorithm outputs the state covariance matrices $P_{k|N}$ and the smoothed estimates for the state vector $x_{k|N}$ that correspond to the trajectories of the model states as well as to the time courses of the model parameters.

\subsubsection*{Expectation-Maximization Algorithm for AKS Initial Parameters}
 In a Kalman Smoother algorithm, appropriate initial values for $x_0$,~$P_0$,~$Q$ and $R$ need be provided. In order to provide optimal starting conditions, an Expectation-Maximization (EM) algorithm can be applied. The expectation step (E-step) consists of the AKS algorithms. Based on the smoothed estimates $x_{k\vert N}$ and $P_{k\vert N}$, maximum-likelihood estimation is performed for $Q$ and $R$, wherefore update equations for these covariance matrices are obtained by maximizing the following target function \cite{dreano2017estimating} with respect to $Q$ and $R$:
 \begin{align}
 \begin{split}
  L = &-\frac{1}{2}\ln\lvert P_0 \rvert- \frac{1}{2} \text{Tr}\left[P_0^{-1} P_{0|N}\right]
 -\frac{N}{2}\ln\lvert Q\rvert - \frac{1}{2}\text{Tr}\left[Q^{-1}\sum_{k=1}^{N}\Sigma_{k}\right]\\
 &-\frac{N}{2}\ln\lvert R\rvert - \frac{1}{2}\text{Tr}\left[R^{-1}\sum_{k=1}^{N}\Omega_{k}\right]\\
\text{where} \qquad \Sigma_k = &P_{k|N}+ (x_{k|N}-b(x_{k-1|N}))\cdot(x_{k|N}-b(x_{k-1|N}))^\text{T}\\
 &+ F_k \cdot P_{k-1|N}\cdot F_k^\text{T}- P_{k-1,k|N} \cdot F_k^\text{T}- F_k\cdot P_{k-1,k|N}^\text{T}\\
  \Omega_{k} = &({y}_k-G\cdot {x}_{k|N})\cdot ({y}_k-G\cdot\textbf{x}_{k|N})^\text{T} + G \cdot P_{k|N} \cdot G^\text{T}
 \end{split}
 \end{align}
It can be analytically shown that the maximum is reached for
  \begin{align}
  	Q_\text{new} = \frac{1}{N} \sum_{k=1}^{N} \Sigma_{k} \qquad \text{and} \qquad	R_\text{new} = \frac{1}{N} \sum_{k=1}^{N} \Omega_{k}\,,
  \end{align}
serving as input for the next EM-iteration \cite{dreano2017estimating}. In addition, the values for $x_{0\vert N}$ and $P_{0\vert N}$ obtained from the AKS of the previous iteration are taken as initial values $x_0$ and $P_0$ of the next iteration corresponding to the maximization step (M-step) of the EM algorithm. These steps are repeated until the convergence criterium reaches a certain level $\alpha$
  \begin{align}
  	\frac{\left|\left(\underset{i}{\sum}Q_i+\underset{i}{\sum}R_i+\underset{i}{\sum}x_{0,i}+\underset{i}{\sum}P_{0,i}\right)-\left(\underset{i}{\sum}Q_{\text{new},i}+\underset{i}{\sum}R_{\text{new},i}+\underset{i}{\sum}x_{0|N,i}+\underset{i}{\sum}P_{0|N,i}\right)\right|}{\left|\left(\underset{i}{\sum}Q_i+\underset{i}{\sum}R_i+\underset{i}{\sum}x_{0,i}+\underset{i}{\sum}P_{0,i}\right)\right|} < \alpha
  \end{align}
with $\alpha$ denoting the relative tolerance for the difference between successive estimates of the hyperparameters $Q$, $R$, $x_0$ and $P_0$. 

Within the combined application with the AKS, an initial guess for these hyperparameters needs to be provided to the EM algorithm. 
In the presented analyses, the choice of these values is not critical for the final results.

\subsubsection*{From the time-continuous ODE system to its time-discrete state space formulation}

Although within this paper, we apply the presented Kalman Smoother method only to epidemic models, it is generally applicable to partially observed non-linear ODE systems, as is shortly outlined in the following. Let us consider the following time-continuous dynamic ODE model
\begin{align}
    \dot x &= f(x,p)\,. \label{eq:eq17}
\end{align}
Here, the time derivative of the state variables $x$ is given as a function $f$ depending on the states vector itself and the parameter vector $p$. The initial conditions of the ODE system can be interpreted as additional parameters in $p$. 

For the time-discrete formalism, the ODE's right-hand-side $f(x,p)$ is interpreted as an update equation for each discrete time step from $x_{k-1}$ to $x_k$, such that Equation~\eqref{eq:eq17} turns into

\begin{equation}
\begin{pmatrix} x_{k} \\ p_{k} \end{pmatrix} = 
\begin{pmatrix} x_{k-1} \\ p_{k-1} \end{pmatrix} +
\begin{pmatrix} f(x_{k-1},p_{k-1}) \\ 0 \end{pmatrix} +\epsilon_{k-1}\,. \label{eq:eq18}
\end{equation}
The parameters $p_k$ are treated as additional states in the state space which are not updated according to the ODE model structure, reflected by the zero term in the equation. 
Instead, the time-dependence originates only from the additive noise term $\epsilon_{k-1}$ which is simultaneously estimated by the Kalman Smoother in each step.

It turns out that this additional degree of freedom enables our Kalman Smoother-based approach to adapt these parameters according to the available data. Thus, by construction it achieves a data-based estimation of time-dependent parameters in the ODE system without any restriction on the parameter trajectory.

\subsection{SIRD-AKS method for estimating the effective reproduction number $\mathcal{R}_t$}

Based on its general derivation in Equation~\eqref{eq:eq18}, the time-discrete state-space formulation of the SIRD model from Equation~\eqref{SIRD_repar} reads
\begin{align}
x_{k} = \begin{pmatrix} S_{k} \\ I_{k} \\ R_{k} \\ D_{k} \\ \mathcal{R}_k \\ \gamma_k \\ \theta_k \end{pmatrix} =
b(x_{k-1})+\epsilon_{k-1} =
\begin{pmatrix} S_{k-1} \\ I_{k-1} \\ R_{k-1} \\ D_{k-1} \\ \mathcal{R}_{k-1} \\ \gamma_{k-1} \\ \theta_{k-1} \end{pmatrix} +
\begin{pmatrix} -\mathcal{R}_{k-1} \cdot (\gamma_{k-1} + \theta_{k-1}) \cdot I_{k-1} \\
 (\mathcal{R}_{k-1}-1) \cdot (\gamma_{k-1} + \theta_{k-1}) \cdot I_{k-1} \\ 
 \gamma_{k-1} \cdot I_{k-1} \\ \theta_{k-1} \cdot I_{k-1} \\
 0 \\ 0 \\ 0 \end{pmatrix}+\epsilon_{k-1} \,,
\end{align} 
where the first four rows correspond to the four states of the time-continuous ODE, and $\epsilon_{k-1}$ is the process noise as introduced in Equation \eqref{state_space}. 
The desired time-dependence of the three SIRD model parameters $\mathcal{R}_k$, $\gamma_k$ and $\theta_k$ is introduced for the AKS method by treating the model parameters as additional states in the state-space formulation which is henceforth called SIRD-AKS.
The estimates from the SIRD-AKS thus yield time courses for the model states $S_k, I_k, R_k$ and $D_k$ as well as for the parameters $\mathcal{R}_k$, $\gamma_k$ and $\theta_k$.

In a real-world scenario of an outbreak, there is typically no data available for the value of the currently infectious individuals in state $I$ per day. Instead, the so-called incidence, i.e.\ the number of newly infected cases per time period, is better accessible and thus usually reported. In the SIRD model, these incidence numbers correspond to the influx in the infectious state $v_{I,k} = \mathcal{R}_{k} \cdot (\gamma_{k} + \theta_{k}) \cdot I_{k}$ between time step $k$ and $k+1$. Analogously, published data for the number of newly recovered $v_{R,k}=\gamma_k\cdot I_k$ and newly dead $v_{D,k}=\theta_k \cdot I_k$ is compared to the respective fluxes. Thus, the fluxes $v_{I,k}$, $v_{R,k}$ and $v_{D,k}$ serve as observables $y_k$ for the SIRD-AKS method. However, the observation matrix $H$ in Equation~\eqref{state_space} only allows for a linear relationship between observables and states. In order to be able to map the data to the appropriate variables of the state-space for the SIRD-AKS method, again the state-space formulation needs to be expanded by three additional states that correspond to the three fluxes, yielding
\begin{align}
x_{k} &= \begin{pmatrix} S_{k} \\ I_{k} \\ R_{k} \\ D_{k} \\ \label{eq_SIRD_AKS}
\mathcal{R}_k \\ \gamma_k \\ \theta_k \\
v_{I,k}\\ v_{R,k}\\ v_{D,k} \end{pmatrix} = b(x_{k-1})+\epsilon_{k-1} =
\begin{pmatrix} S_{k-1} \\ I_{k-1} \\ R_{k-1} \\ D_{k-1} \\
\mathcal{R}_{k-1} \\ \gamma_{k-1} \\ \theta_{k-1} \\
0 \\ 0 \\ 0 \end{pmatrix} +
\begin{pmatrix} -v_{I,k-1} \\ v_{I,k-1} - v_{R,k-1} - v_{D,k-1} \\ v_{R,k-1} \\ v_{D,k-1} \\
0 \\ 0 \\ 0 \\ \mathcal{R}_{k-1} \cdot (\gamma_{k-1} + \theta_{k-1}) \cdot I_{k-1} \\
 \gamma_{k-1} \cdot I_{k-1}\\ \theta_{k-1} \cdot I_{k-1} 
 \end{pmatrix}+\epsilon_{k-1}
 \end{align}
and observation matrix
\begin{align}H &= \begin{pmatrix}
 & 1 & 0 & 0 \\
\mathbb{0}_{7\times 3} & 0 & 1 & 0 \\
 & 0 & 0 & 1 
\end{pmatrix}\,.
\end{align}
that links the incidence data to the respective fluxes fluxes $v_{I,k}$, $v_{R,k}$ and $v_{D,k}$.
Based on incidence data, the SIRD-AKS method is applied to the state-space formulation~Equation~\eqref{eq_SIRD_AKS} and yields estimates for the time courses of the state-space vector $x_k$. Its entries are then interpreted as discretized time series of the model states, time-dependent parameters and fluxes, i.e.\ incidence numbers.
Since all variables are strictly positive, all analyses are performed on logarithmic scale for $x_k$, which is beneficial for numerical stability.

\subsection{Incidence-based reproduction number calculation method}\label{Incidence}

In contrast to the formulation of the effective reproduction number $\mathcal R_t$ as a time-dependent parameter in the reparameterized SIRD model and its estimation using e.g.\ the SIRD-AKS method, the reproduction number can also be calculated directly from incidence data \cite{cori2013new}.
In its original formulation in \cite{cori2013new}, this incidence-based method requires not only incidence data, but also the empirical serial interval distribution, i.e.\ information about the time period between illness onset in an infected case and illness onset in a consecutive case.
Since reliable data on this was not available in the beginning of the SARS-Cov-2 pandemic, the German federal center for disease control and prevention, i.e. the Robert Koch Institute (RKI), chose to use the incidence-based method from \cite{cori2013new} and assumed a Dirac $\delta$-distribution at $s = 4$ days for the serial interval distribution \cite{hamouda2020schatzung}.
This assumption of one single applicable value for the  serial interval time $s$ leads to an easily interpretable calculation rule for the effective reproduction number as the ratio 
\begin{align}
\mathcal R_{t}^s &= \frac{n_{I,t}}{n_{I,{t-s}}} \label{r}
\end{align}
between the number of newly infected individuals $n_I$ at time point $t$ and at time point $t-s$.
To account for fluctuations in the reported incidence data such as week-day dependencies, a moving average over $\tau$ days is considered that leads to 
\begin{align}
\mathcal R_{t}^{s,\tau} 
&= \frac{\frac{1}{\tau}\sum_{i=t-\tau +1}^{t} n_{I,i} }{\frac{1}{\tau}\sum_{i=t-\tau +1}^{t}n_{I,{i-s}}}= \frac{\bar{n}_{I,i}^\tau}{\bar{n}_{I,{i-s}}^\tau}. \label{eq_R_tau}
\end{align}
It should be noted, that two different $\mathcal R_{t}^{s,\tau}$-values were published on a daily basis by the RKI, both with the described assumption of the $/delta$-distributed serial interval at $s=4$ days: (i) A so-called sensitive 4-day $\mathcal R_{t}^{4,4}$ value, and (ii) a more stable 7-day $\mathcal R_{t}^{4,7}$, depending on the averaging window length.

The advantage of this incidence-based method is that it does not require the explicit formulation of a compartment model, its numerical evaluation or any estimation of parameters. 
Furthermore, a certain value of for example $\mathcal R_t = 2$ becomes easily interpretable and verifiable also for non-experts, since it corresponds to a doubling in the numbers of newly infected individuals within $s = 4$ days.
As a consequence, the daily updated value of $\mathcal R_{t}$ calculated from this incidence-method was used by the RKI to inform the public as well as political decision makers about the status of the pandemic situation and thus had a decent influence on the installed non-pharmaceutical interventions \cite{hamouda2020schatzung}.

While being very simple to calculate and easy to interprete, the method has the disadvantage that the $R_t$ value and its interpretation strongly depends on the appropriateness of the chosen $\delta$-distributed serial interval, as it is directly associated with the time period for the doubling times. In contrast, the presented SIRD-AKS method does not require any assumptions on the serial interval distribution.   

\subsection{Simulation setting} 

To show the performance of the presented AKS method for the estimation of time-dependent parameters in nonlinear ODE models, we first applied the algorithm to simulated data sets from various scenarios. In each simulation scenario, a different combination of model structure and values of the time-constant parameters or time courses of the time-dependent parameters were chosen. The ODE system was then solved for given initial conditions by using the \texttt{R} package \texttt{deSolve} \cite{soetart2010deSolve}.
In order to generate realistic incidence data for the respective quantities, the fluxes $v_I$, $v_R$ and $v_D$ of the ODE system were discretized and relative Gaussian noise with zero mean and given standard deviation was added.
In the initial examples, different parameter time courses and noise levels were chosen in an SIRD-model for data generation, followed by scenarios that consider data sets generated by non-SIRD models. Ultimately, the method was applied to officially reported incidence data for SARS-CoV-2 in Germany.

Note that all noisy incidence data sets were analyzed by the presented SIRD-AKS method, i.e.\ based on the reparameterized SIRD model from Equation\ \eqref{eq_SIRD_AKS}, regardless of the chosen model structure for data generation.
Further, no information about true values, time-dependency or function class of the model parameters were provided to the SIRD-AKS algorithm, but only the respective incidence data sets.
Then, the resulting estimates of the SIRD-AKS algorithm for model states, fluxes and parameter time-courses were compared to the respective true quantities derived from the solutions of the data generating ODE models.
In particular, the accordance of the estimated time-dependent effective reproduction number $\mathcal{R}_t$ and its true time course was analyzed.

\section{Results}\label{results}

\subsection*{AKS performance for multiple time-varying SIRD model parameters} 

\begin{figure}
	\centering
	\includegraphics[width=1\textwidth]{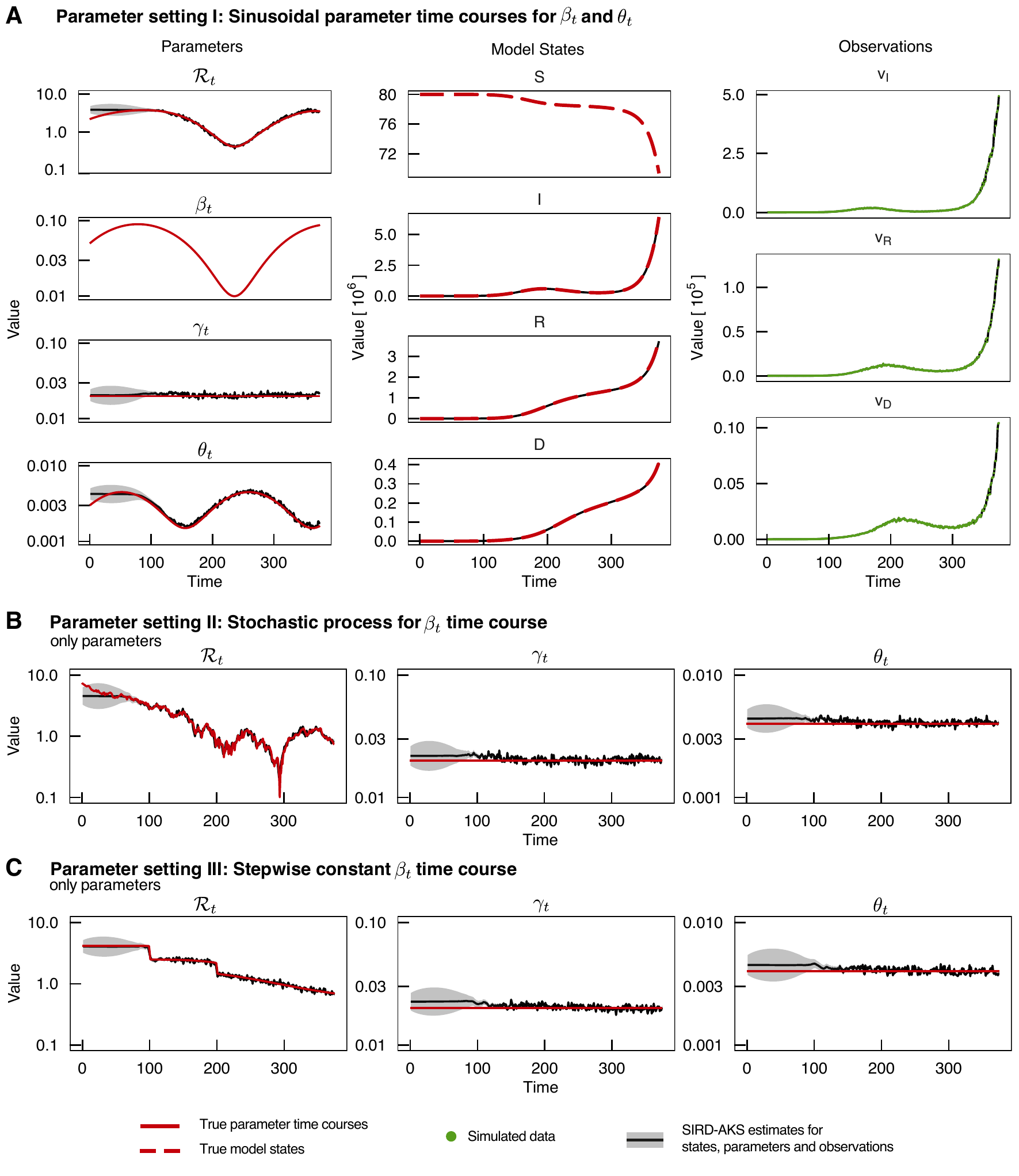} 
	\caption{\footnotesize Estimation of multiple time-dependent parameters. \textbf{A:}~The classical SIRD model was simulated with a time-constant parameter $\gamma$ and time-dependent parameters $\beta_t$ and $\theta_t$. From these parameters, the effective reproduction number $\mathcal{R}_t$ was calculated via Equation~\eqref{R}. True parameter time courses and states are depicted as red solid lines and red dashed lines, respectively. Thereof, data was simulated at 375 equidistant time points for the observed fluxes $v_D$, $v_I$ and $v_R$ as indicated by green dots. The SIRD-AKS is based on the reparameterized SIRD model and thus does not yield estimates for $\beta_t$ or $S$ but instead directly for $\mathcal{R}_t$.
	The estimates of the SIRD-AKS are shown as black solid lines with gray bands for their uncertainties. Results are shown on logarithmic scale. The results of the the SIRD-AKS algorithm are able to describe the data and are in good agreement with the true parameter time courses, both for the time-constant and time-dependent parameters. Note that the SIRD-AKS estimates for the states overlap with the true time courses of the model states.
	\textbf{B-C:}~ In different parameter settings, simulation was performed with time-constant parameters $\gamma$ and $\theta$, whereas $\beta_t$ was either chosen as a stochastic process (\textbf{B}), or as a step function (\textbf{C}), respectively. The corresponding values for $\mathcal{R}_t$ were computed using Equation~\eqref{R}. In both cases, the SIRD-AKS results retrieve the true parameter time courses adequately.}\label{fig2_concept}
\end{figure}

As a first simulation setting, the classical SIRD model from Equation~\eqref{SIRD} with the initial values $S_0=10^8$, $I_0=1$ was used for data generation.
Furthermore, three different parameter settings for the three model parameters $\beta_t,~\theta_t$ and $\gamma_t$ were simulated.
In the first parameter setting shown in Figure~\ref{fig2_concept}A, time-varying parameters were chosen as $\beta_t=0.05+0.04\cdot \sin{(t/50)}$ and $\theta_t=0.003+0.0015 \cdot \sin{(t/33)}$, while $\gamma=0.02$ was chosen to be constant over time.
The resulting dynamics for $\mathcal{R}_t$ reflect the combination of the sinusoidal input $\beta_t$ and the decrease of susceptibles $S$, c.f.~Equation~\eqref{R}.
To generate data sets, the fluxes $v_I=\beta_t \frac{S I}{N_0}$, $v_D=\theta_t I$ and $v_R=\gamma_t I$ between the model's states were evaluated at 375 equidistant time points, mimicking approximately one year of daily observations.
Gaussian noise with a relative error of $5\%$ was added to the three time series of observables, shown as green dots in Figure~\ref{fig2_concept}.

Next, the noisy data was provided to the SIRD-AKS in order to estimate model states and time-dependent parameters. For this, the EM-algorithm part of the SIRD-AKS was initialized with $I_0=100,~R_0=0.1,~D_0=0.1,~v_{I,0}=1,~v_{R,0}=0.1,~v_{D,0}=0.1,~\gamma_0=\log(0.1),~\theta_0 =\log(0.01),~\mathcal{R}_{t,0}=\log(6)$ and the convergence threshold was chosen to be $\alpha = 0.001$. The estimation results are shown as black lines in Figure~\ref{fig2_concept}A, where the grey error bands of the estimates correspond to the estimated covariance matrices $P_{k \vert k}$. Note that the time-dependent parameters are shown on logarithmic scale throughout this work. The SIRD-AKS is able to simultaneously capture the dynamics of the simulated values for parameters, states and observations. In particular, the time-dependent parameters $\gamma_t$, $\mathcal{R}_t$ and $\theta_t$ were obtained without providing any information on their function class or dynamical behavior. It can be observed that the grey error bands for the estimated parameters in Figure~\ref{fig2_concept} are larger in the beginning of the time series, where also the parameter estimates slightly deviate from the true values due to missing dynamics in the data. In contrast, with increasing dynamic activity in the data, the error bands become smaller and the estimates coincide with the true values.

To analyze whether the SIRD-AKS can also handle different time-dependencies of the model parameters, for example rapidly changing parameters including discontinuous functions over time, two further parameter settings were addressed as shown in Figures~\ref{fig2_concept}B and C. For this, the SIRD model was evaluated with time-constant parameters $\gamma_t$ and $\theta_t$ and a time-varying $\beta_t$. 
In parameter setting II, the time course of $\beta_t$ was chosen as an autoregressive process of order $p=1$ (AR[1]) with an added exponential decay. The autoregressive process is given by $x_i = 0.0022+0.95\,x_{i-1}+w_i$ with noise term $w_i \sim \mathcal{N} (0, 0.004)$.
Furthermore, a step function with three constant levels was utilized in parameter setting III. Data was simulated (not shown) and the SIRD-AKS algorithm was applied analogously to the previous setting I with sinusoidal $\beta_t$. For all parameter settings I, II and III, the SIRD-AKS results adequately capture the dynamics of the three time-dependent parameters, demonstrating a broad applicability of the method to different function classes of time-dependent parameter time courses.

\subsection*{Performance for high noise levels} 

\begin{figure}
	\centering
	\includegraphics[width=\textwidth]{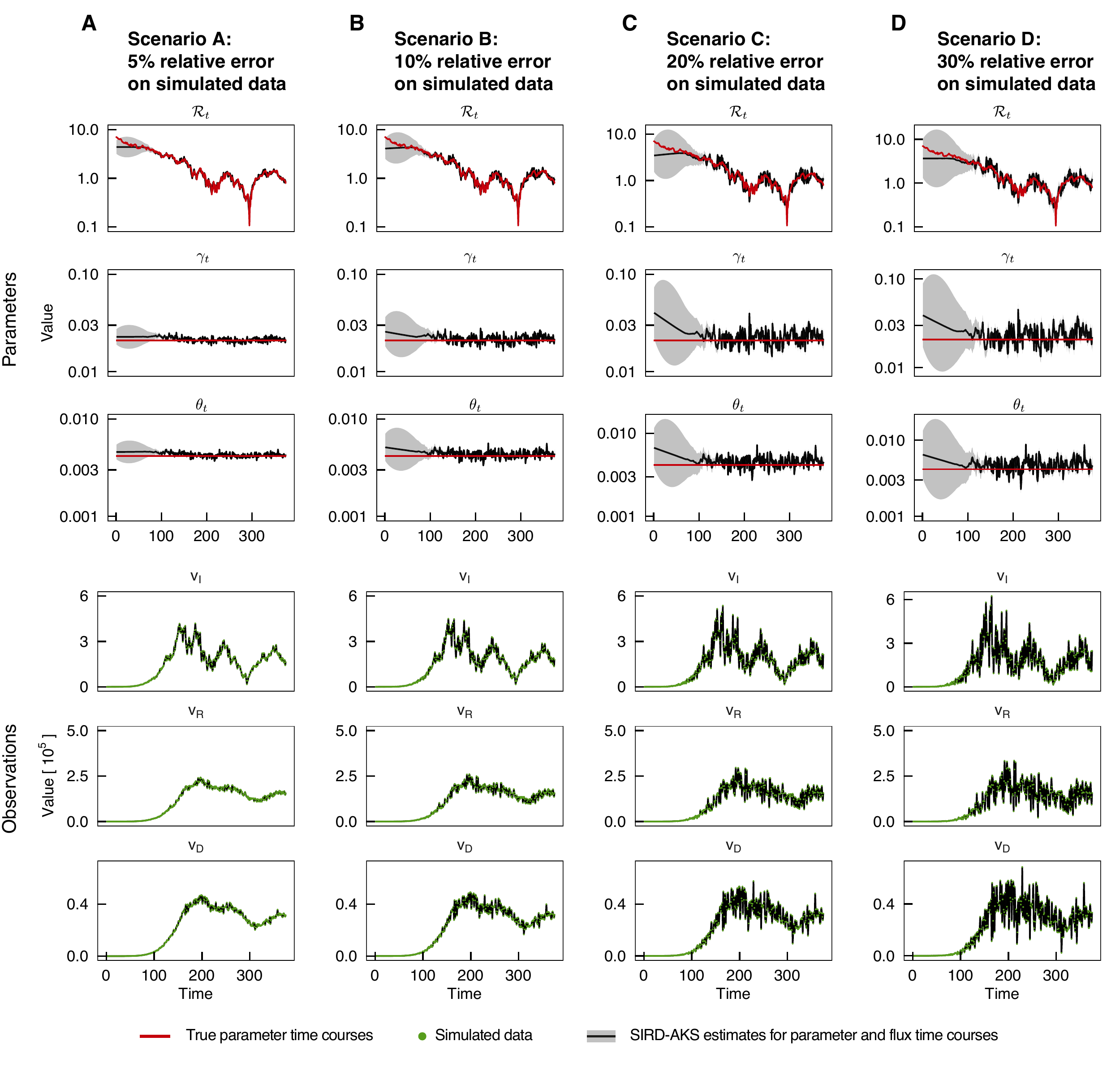} 
	\caption{\footnotesize Performance of SIRD-AKS is robust towards high noise levels in the data. \textbf{A-D:}~The SIRD model was simulated with a time-varying $\beta_t$ and time-constant $\gamma_t$ and $\theta_t$ as shown in red. The parameters were chosen as in parameter setting II in Figure~\ref{fig2_concept}. The data was simulated using increasing noise levels from scenario \textbf{A} to \textbf{D} as shown as green dots for the observed fluxes $v_D$, $v_I$ and $v_R$. The SIRD-AKS estimates are shown as black lines with grey error bands for their uncertainties and are compared to the corresponding true time courses. Again, instead of parameter $\beta_t$ from the data generating SIRD model, $\mathcal{R}_t$ was calculated from the true parameter time courses and compared to the SIRD-AKS result. 
	While the precision of the SIRD-AKS estimates decreases with higher noise levels, the average dynamics of the parameters time courses are satisfactorily met.}\label{fig3_noise}
\end{figure}

To investigate how robust the SIRD-AKS method performs with respect to possibly high observation noise, we again considered parameter setting II with the stochastic process for $\mathcal R_t$. In parallel, four data sets were simulated with different noise levels as shown in Figures~\ref{fig3_noise}A-D. The parameter time courses and initial values for the states were chosen to be the same for all these scenarios. This is reflected by the true value of parameter $\mathcal{R}_t$ that shows the same trajectory in all four scenarios, represented by the red line in Figure~\ref{fig3_noise}. In contrast, the simulated data for the observed fluxes expectedly shows higher variations for higher noise levels, shown in green in Figure~\ref{fig3_noise}.

The SIRD-AKS method was applied to all four scenarios and its estimates are shown as black lines in Figure~\ref{fig3_noise}. Firstly, it can be seen that the observations are adequately described by the SIRD-AKS which follows the data points also when they rapidly vary over time as in Figure~\ref{fig3_noise}D. Secondly, the correct trajectory of the time-dependent parameter $\mathcal{R}_t$ as well as the correct levels of the time-constant parameters $\gamma$ and $\theta$ were obtained in all four scenarios. However, the higher the noise level, the more it becomes evident that the SIRD-AKS estimates fluctuate around the true values of the parameters, while the average form of the parameter time course is still in accordance with the time course for the true value of $\mathcal{R}_t$. Taken together, these results show that the SIRD-AKS results for the estimates of the time-dependent parameters remain correct, even for relatively high noise levels in the data.

\subsection*{Influence of potential model misspecifications}

Up to this point, we have only shown applications of the SIRD-AKS method to simulated data from the same SIRD model structure.
In more realistic applications, however, the true model structure is not known, or as in all real-world scenarios, the chosen model structure is only an appropriate simplification of the real underlying biological process.
Therefor, we next investigated the performance of the SIRD-AKS method in the light of model misspecifications.
To this aim, data was generated from non-SIRD model structures, but analyzed by the AKS method using the \emph{inappropriate} SIRD model structure. We tried to reconstruct the time courses of the effective reproduction number in two examples of such intentionally misspecified model structures for the SIRD-AKS method.

\begin{figure}
	\centering
	\includegraphics[width=1\textwidth]{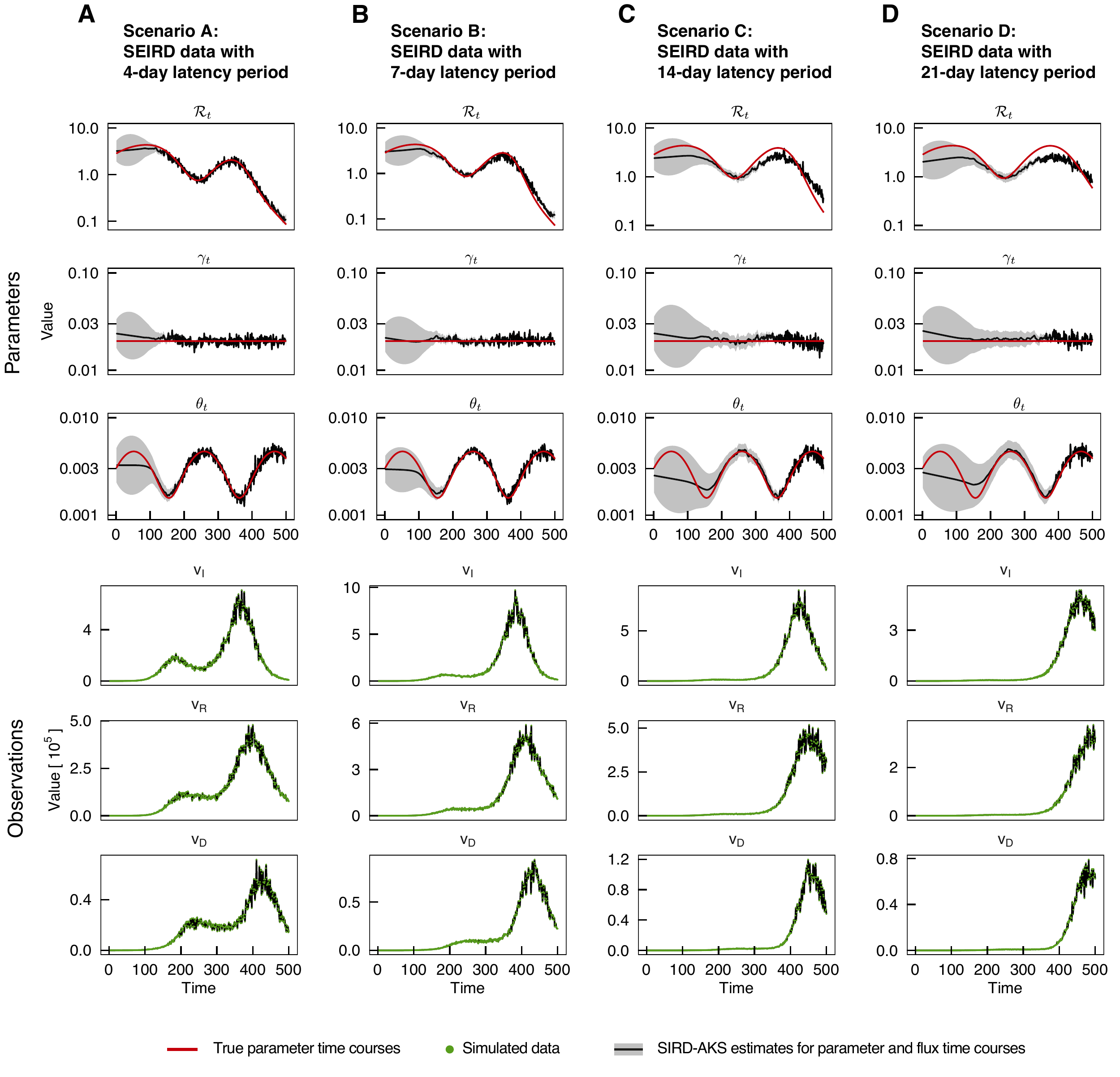} 
	\caption{\footnotesize AKS shows robustness to model misspecifications. \textbf{A-D:}~The SEIRD model from Equation~\eqref{SEIRD} was simulated with time-varying parameters $\beta_t$ and $\theta_t$ and for different latency periods increasing from $\delta^{-1}=4d$ (\textbf{A}) to $\delta^{-1}=21d$ (\textbf{D}). Calculated time courses for $\mathcal{R}_t$ depend on the latency period parameter and thus changes in the four scenarios. Simulated data was generated from the SEIRD model fluxes with $5\,\%$ relative error. The data was analysed by the SIRD-AKS algorithm, which assumes the reparameterized SIRD model as truth and, in particular, does not contain an Exposed state $E$. Despite the model misspecification between the data generating process and the analyzing model, the SIRD-AKS estimates for the time-dependent parameters are accurately met for shorter latency periods in Scenario \textbf{A-B}. For larger deviations from the assumed SIRD model structure, the AKS results are slightly biased towards values of one (\textbf{C-D}), while the overall dynamics is retrieved sufficiently.}\label{fig4_misspecification}
\end{figure}

First, the so-called SEIRD model was considered for data generation. It constitutes an extension of the SIRD model augmented by a model state of Exposed (E) corresponding to individuals which are infected but not yet infectious. This extension is commonly considered when the disease progression underlies an incubation period. The ODE system of the SEIRD model reads
\begin{align}
\begin{split}
\dot{S} &= -\beta\cdot\frac{S\cdot I}{N_0}  \label{SEIRD} \\
\dot{E} &= \beta\cdot\frac{S\cdot I}{N_0} - \delta \cdot E \\
\dot{I} &= \delta \cdot E -\gamma \cdot I-\theta \cdot I \\
\dot{R} &= \gamma \cdot I \\
\dot{D} &= \theta \cdot I \,,
\end{split}
\end{align}
where the reciprocal of the additional parameter~$\delta^{-1}$ is called latency period. Note that the limiting case of the SEIRD model for $\delta^{-1}\rightarrow 0$ is the SIRD model, while for higher values of $\delta^{-1}$ the difference between the two models increases and distinct dynamics are expected. To illustrate this transition, four different analyses with latency periods varying from 4 to 21 days were performed.

The SEIRD model Equation~\eqref{SEIRD} was simulated for data generation with time-dependent parameters $\beta_t$ and $\theta_t$ as well as time-constant parameter $\gamma_t$, following the previous parameter setting I. Again, the true values for the reproduction number~$\mathcal R_t$ were calculated from the model parameters for the SEIRD model \cite{van2017reproduction}. 
To mimic a realistic application of the presented method in an epidemic scenario, where the true underlying process is not fully known, data from the SEIRD model was analyzed by the SIRD-AKS and only the daily numbers of newly infected $v_I$, newly recovered $v_R$ and newly dead $v_D$ were provided as observations to the algorithm. 
In consequence, the SIRD-AKS is able to estimate the SIRD model states and parameters, but cannot infer the transition rate $\delta$ or the exposed state $E$.

The resulting time courses of the SIRD-AKS estimates are shown in the upper part of Figure~\ref{fig4_misspecification} with increasing values for the latency period $\delta^{-1}$ from panel A to D. 
Comparing the four simulation scenarios, the different latency periods are reflected both in different dynamics of the observables and in the time courses of the reproduction number $\mathcal{R}_t$. As expected, the peak in the observed infections appears later for higher latency periods corresponding to longer times that individuals remain in the exposed state $E$. 

For all scenarios, the SIRD-AKS was applied to the SEIRD data sets, as shown in black in Figures~\ref{fig4_misspecification}A-D. It can be seen that the time courses of the parameters $\gamma_t$ and $\theta_t$ are throughout accurately obtained. Interestingly, the estimates for the effective reproduction number $\mathcal R_t$ are also very close to the true values for low to medium latency periods. In contrast, for higher latency periods, the estimates for $\mathcal R_t$ are slightly shifted to later time points and also biased towards the characteristic value of $\mathcal R_t=1$.
In total, this shows that the AKS method is able to cope with moderate deviations in the model structure, while more extreme model misspecifications, e.g.~latency periods of three weeks, can noticeably influence the result of the estimated time-dependent parameters.
Especially, for COVID-19 such a model misspecification does not negatively impact the SIRD-AKS estimation results, since typically a latency period of 5-6 days was reported \cite{xin2021latent}.

\begin{figure}
\centering
    \includegraphics[width=1\textwidth]{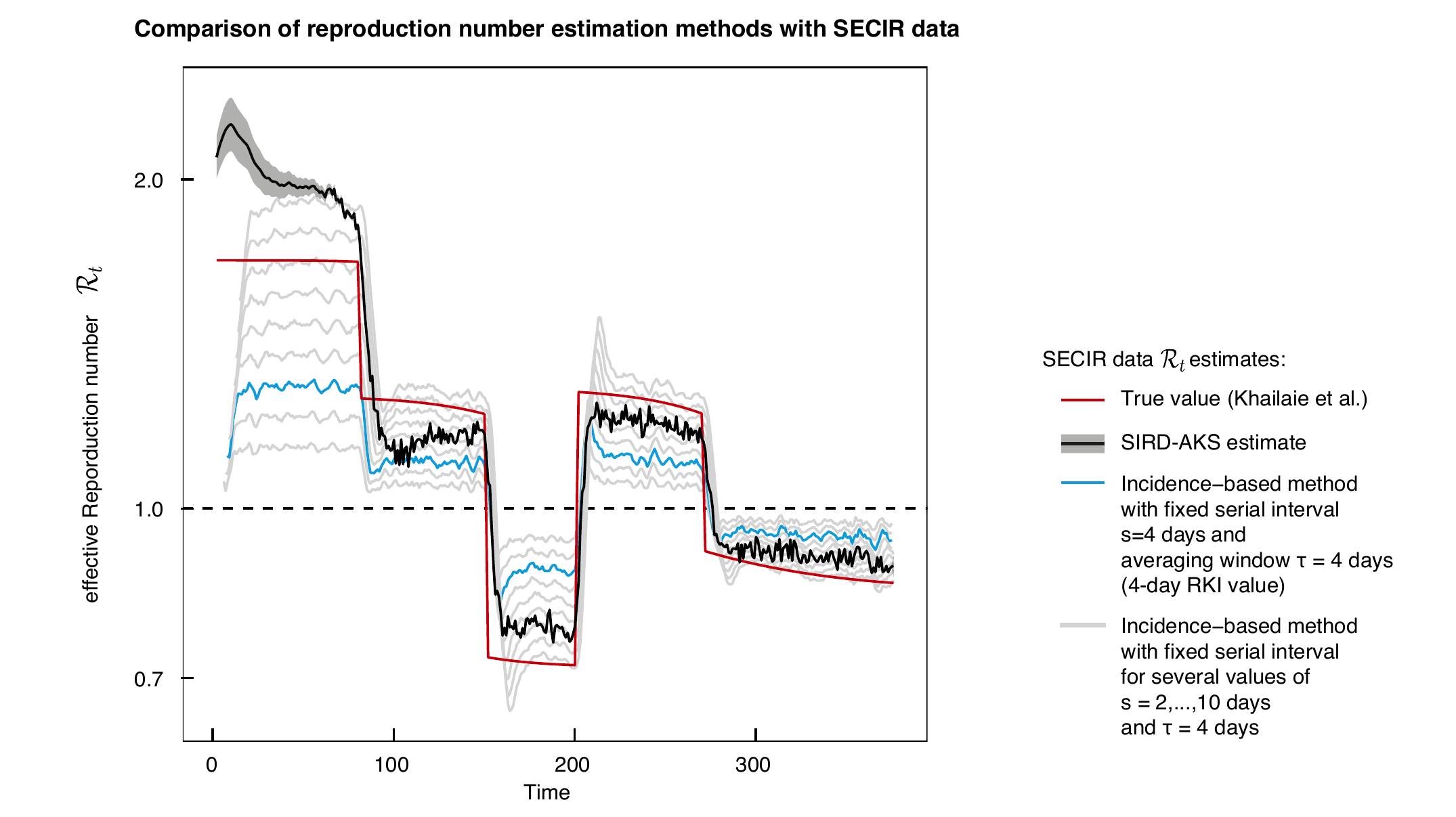}
	\centering
	\caption{\footnotesize Comparison of SIRD-AKS estimation and incidence-method with fixed serial interval time. Simulated data was generated from the SECIR model \cite{khailaie2021development} with five-step function as time-dependent input for parameter $R_1$. The corresponding true time course for $\mathcal R_t$ is depicted as red line. The SIRD-AKS estimate as shown in black roughly reassembles the true SECIR reproduction number. The $\mathcal{R}_t^{s,\tau=4}$ value calculated by the incidence-based method from Equation \eqref{eq_R_tau} with fixed $s=4$ is shown in blue. It is biased towards $\mathcal R_t=1$ (dashed line) and is not able to correctly estimate the true time course. For different choices of the fixed serial interval time hyperparameter $s$, the incidence-based method may yield better results for $\mathcal{R}_t^{s,4}$, as shown by the grey lines for different values of $s = 2, ... , 10$.}\label{fig5_SECIR}
\end{figure}

In order to formulate a model-misspecification problem that is closer to a real-world scenario, we additionally utilized a larger and more complicated model structure that covers a multitude of additional features. For this, we employed the so-called SECIR model that consists of sixteen model states and has been developed for a detailed description of the SARS-CoV-2 epidemic in Germany in 2020, including for example several infected carrier states as well as diverse hospitalization stages \cite{khailaie2021development}. This model was used for data generation by using parameters values within the allowed parameter ranges as in the original publication. The only deviation is that we replaced the originally estimated time course of the transmission parameter $R_1$ by a simple five-step function. This parameter mainly influences the time course of the true value of the effective reproduction number $\mathcal R_t$ within the SECIR model, which was calculated according to the respective equation in the original publication and is shown as a reference in red in Figure~\ref{fig5_SECIR}. Within the chosen parameter set for the simulation, the resulting true $\mathcal R_t$ time course covers a broad range of values, in particular below and above the characteristic value of $\mathcal R_t = 1$.
The generated data sets for the newly infected $v_I$, newly recovered $v_R$ and newly dead $v_D$ were again provided as observations of the SIRD-AKS analysis. The estimated time course of the effective reproduction number $\mathcal R_t$ from the SIRD-AKS with SECIR data is shown as black line in Figure~\ref{fig5_SECIR}. 
Except for the level of the first plateau until $t=80$, the $\mathcal R_t$ value from the SIRD-AKS method shows a similar shape as the true $\mathcal R_t$ time course and yields almost correct levels on the other plateaus. However, the estimated $\mathcal R_t$ time course is slightly biased to $\mathcal R_t=1$ after $t = 80$ days.

For a comparison of the different methods, Figure~\ref{fig5_SECIR} also shows the results for the reproduction number $\mathcal{R}_t^{s,\tau}$ calculated by the incidence-based method from Equation~\eqref{eq_R_tau}
 
for different values of the fixed serial interval time $s=2, ..., 10$ days in unit steps, and with averaging windows $\tau=4$ days. It can be clearly seen that the choice of the fixed serial interval time heavily influences the resulting levels of the reproduction number. 
The Robert Koch Institute (RKI) as German center for disease control provided officially-issued reproduction number values during the SARS-Cov-2 pandemic for Germany based on this method with a chosen fixed serial interval value of $s=4$ days.
The accordingly calculated $\mathcal{R}_t^{4,4}$ value for the SECIR data set is highlighted in blue in Figure~\ref{fig5_SECIR}.
Comparing blue and red line, it can be concluded that at least for the simulations of the SECIR model and the chosen parameterization, a fixed serial interval of $s=4$ days is not an adequate choice. While the RKI-issued value roughly covers the general form of the time course, it is noticeably biased towards the characteristic value of $\mathcal R_t=1$. As a consequence, the rate of increase or decay in newly infected individuals within an ongoing disease outbreak would be definitely underestimated. 

An appropriate choice of the serial-interval distribution or at least an optimal choice of the fixed serial interval time $s$ would solve this issue, as can be seen from the grey lines in Figure~\ref{fig5_SECIR}. However, it should be noted that the selection of the optimal value for $s$ is difficult in real-world applications and requires additional data or assumptions. That is, the simplicity of the calculation and interpretation using the incidence-based method comes at the price of required assumptions for the serial interval times.

Some deviations from the true $\mathcal R_t$ value also appear in the results of the SIRD-AKS method. However, we surmise that these originate from the misspecified model structure. We argue that a clear advantage of the SIRD-AKS method is that it does not rely on additional hyperparameters or distributional assumptions, such as the incidence-based method does on the serial interval time. This renders the SIRD-AKS method to be a \emph{de facto} non-parametric approach for the estimation of the reproduction number and model parameters in general. As shown in this example, the SIRD-AKS method also yields reasonable results, when the internal state space model does not meet the model structure of the data generating process. 
It therefore qualifies for the application to real-world data sets, where the true underlying process is masked and thus the optimal choice of an appropriate fixed serial interval time for the incidence-based method becomes even more difficult.

\subsection*{Application to SARS-CoV-2 data from Germany}

\begin{figure}[H]
    \centering
	\includegraphics[width=\textwidth]{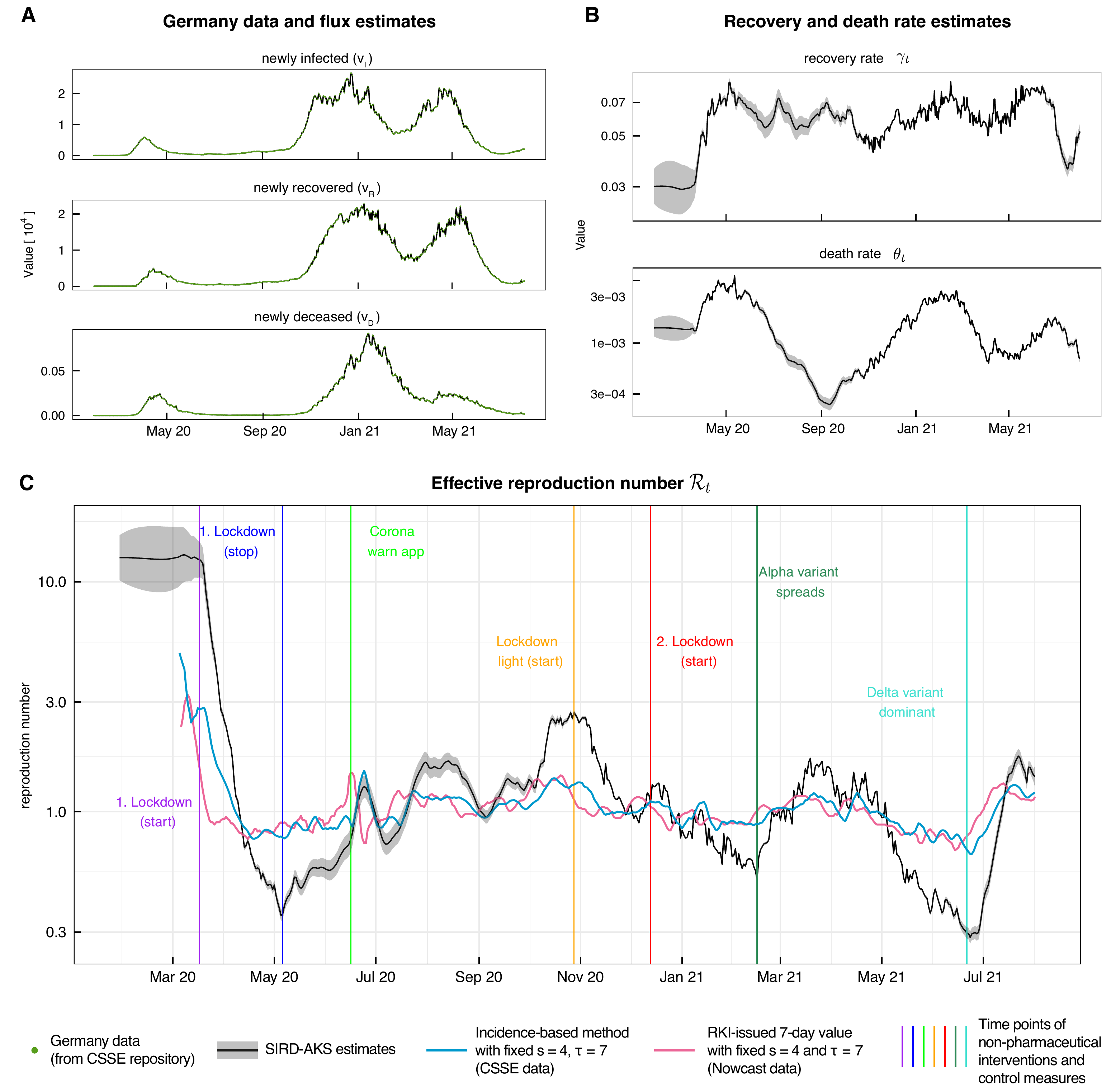} 
	\caption{\footnotesize Application of the SIRD-AKS to SARS-CoV-2 data of Germany from the CSSE repository. \textbf{A:}~Incidence data taken from \cite{dong2020interactive} for March 2020 until August 2021 is shown in green. SIRD-AKS results  for the fluxes of newly infected, newly recovered and newly deceased is shown as black line. \textbf{B:}~After an initial rise, the SIRD-AKS estimated for the  recovery rate $\gamma_t$ stays nearly constant, while the estimated death rate $\theta_t$ shows a dynamic with two peaks at around May-June 2020 and February-March 2021. These time points coincide with periods of high hospitalization states and where mostly older people were infected \cite{staerk2021estimating}. \textbf{C:}~The SIRD-AKS estimates (black) for the reproduction number are displayed in comparison to the seven-day reproduction number based on Nowcast data and published by the RKI in magenta \cite{hamouda2020schatzung} as well as the calculated $\mathcal R^{s,\tau}_t$ incidence-based method in blue that is based on CSSE data. Both incidence-based methods assume a fixed serial time of $s = 4$ days and averaging window of $\tau = 7$ days. Time points of essential control measures are indicated by colored vertical lines \cite{website}.}
	\label{fig6_germany}
\end{figure}

After demonstrating the performance of the SIRD-AKS method for multiple parameter time courses and model misspecifications in simulated data sets, the method was applied to COVID-19 data from Germany for January 2020 until August 2021. The data was taken from the COVID-19 Data Repository maintained by the Center for Systems Science and Engineering (CSSE) at Johns Hopkins University \cite{dong2020interactive}, and is displayed in Figure~\ref{fig6_germany}A as green dots. Data pre-processing was performed with a moving average over seven days in order to reduce week-day effects and reporting delays in the raw data. 

Figure~\ref{fig6_germany}A shows the analyzed data and estimated time courses of the observed fluxes for the newly infected $v_I$, newly recovered $v_R$ and newly dead $v_D$. The estimated time courses for the recovery rate $\gamma_t$ and death rate $\theta_t$ are shown in Figure~\ref{fig6_germany}B. The fluctuations in the death rate are in line with the time periods when more old people had been reported to get infected, i.e.~from April to June and from November to February 2020 \cite{staerk2021estimating}. This also agrees with previous studies showing that the mortality of COVID-19 is heavily linked to the patients age \cite{yanez2020covid,ho2020older}.

The estimated time-dependent reproduction rate $\mathcal{R}_t$ from the SIRD-AKS method is shown as black line in Figure~\ref{fig6_germany}C. For comparison, two additional $\mathcal{R}_t$ value are provided, which are both based on the previously discussed incidence-based method and assume a fixed serial interval of $s=4$ days and a averaging window of $\tau = 7$ days.
However, they use different data sources. The $\mathcal{R}_t$ value shown as blue line is calculated from the same data from \cite{dong2020interactive} that is also analyzed by the SIRD-AKS method. The other shown value in mangenta is the officially-issued and so-called \emph{seven-day} $\mathcal{R}_t^{4,7}$ value from the RKI \cite{hamouda2020schatzung}.
The difference is that it does not use the raw data from the CSSE repository, but instead applies an additional so-called \emph{Nowcasting} pre-processing method that copes with the redistribution of the case numbers for the elimination of reporting delays \cite{hohle2014bayesian}. Both incidence-based methods show a similar time course for $\mathcal{R}_t$ but with peaks that seem to be shifted by a couple of days, presumably because of the case number redistribution.

When compared to the estimate from the SIRD-AKS method, both incidence-based values again lie closer to $\mathcal{R}_t = 1$. The time course from the SIRD-AKS shows a larger variety of peak heights in both directions and more deviations from $\mathcal{R}_t=1$ on longer time-scales. This becomes even more prominent when the changes in the reproduction numbers are compared to the time points of new non-pharmaceutical interventions and control measures as indicated by vertical lines. While the incidence-based $\mathcal{R}_t$ values likewise decrease drastically at the start of the first nation-wide lockdown in March 2020, the end of the first lockdown is only visible as a trend change of the SIRD-AKS reproduction number, but not in the incidence-based approaches. A similar case occurs for the so-called lockdown light in November 2020 and second nation-wide lockdown in December 2020 \cite{website}. Both incidence-based measures barely show any effect on the reproduction number, while the SIRD-AKS estimate shows a fluctuating, yet undoubtedly decreasing trend until the approximate time point where the SARS-CoV-2 Alpha variant spread in Germany.

\subsection*{Validation of parameter time course estimates in an ODE model}

The presented SIRD-AKS method is based on the transition of the model structure from the time-continuous ODE system to the time-discrete and recursive formulation in the state space. Further, estimates for the time-dependent parameters are in fact augmented states in the state space, interpreted as parameter time courses. We asked whether these parameter time courses produce coherent dynamics when being incorporated into the corresponding ODE system as input functions.

\begin{figure}[H]
    \centering
	\includegraphics[width=\textwidth]{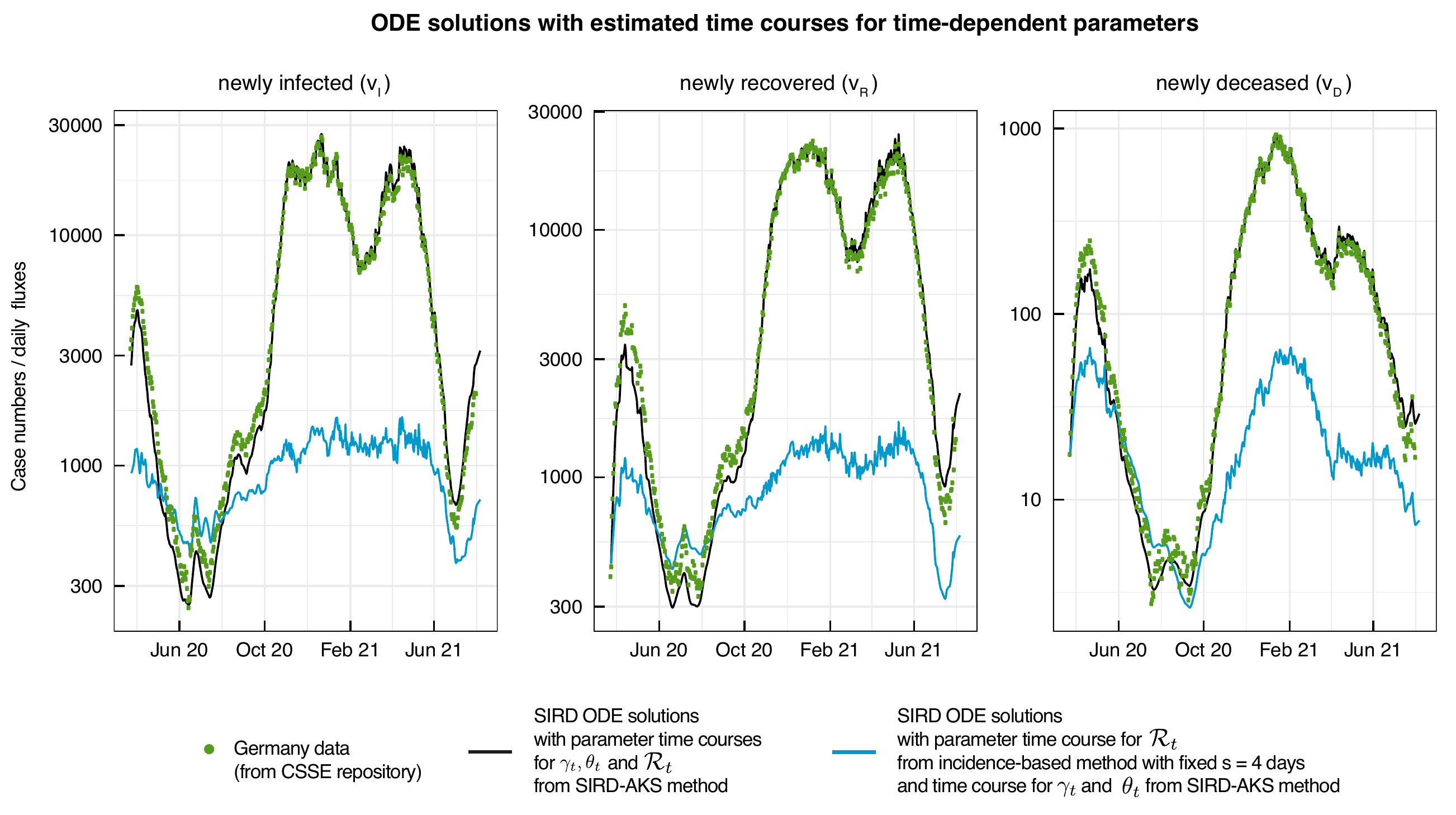} 
	\caption{\footnotesize ODE solutions for the SIRD model with estimated time-dependent parameters. The fluxes for $v_I$, $v_R$ and $v_D$ are calculated from the numerically solved ODEs of the reparameterized SIRD model from Equation~\eqref{SIRD_repar} and are compared to Germany data from Figure~\ref{fig6_germany}A as depicted by green dots. The black lines indicate the ODE solutions when the three estimated time courses for $\gamma_t, \theta_t$ and $\mathcal{R}_r$ from Figure~\ref{fig6_germany}B and C are utilized as time-dependent parameters in the SIRD ODE system. Identical ODE simulations were performed and are shown as blue line, where the $\mathcal R_t$ time course is taken from the incidence-based method with fixed $s = 4$ and $\tau = 7$ (c.f. Figure~\ref{fig6_germany}C blue line). While the SIRS-AKS estimates yield almost perfect ODE trajectories, the incidence-based method underestimates the dynamics of the pandemic in an SIRD model.}
	\label{fig7_ring}
\end{figure}

To perform this consistency check, we plugged in the estimated time courses for the parameters $\mathcal R_t,~\gamma_t$ and $\theta_t$ from Figure~\ref{fig6_germany}B and C into the ODEs of the reparameterized SIRD model in Equation\ \eqref{eq_SIRD_AKS}.
To choose appropriate initial values for the numerical solution of this ODE system, the values of the SIRD-AKS estimates at time point $k=62$, i.e.~March 24, 2020, were chosen. This time point corresponds to the first time point where the estimated uncertainties of the time-dependent parameters become adequately small. It also coincides with the starting point of the large initial decrease in the time course of $\mathcal{R}_t$, c.f.\ Figure~\ref{fig6_germany}C.
Results of the numerical simulation of the time-continuous ODE system with time-dependent parameters are shown in Figure~\ref{fig7_ring} as black line. It is remarkable that all three trajectories are very well in agreement with the flux data for $v_I$, $v_R$ and $v_D$ that went into the previous AKS analysis, shown as green dots in Figure~\ref{fig7_ring}.

This confirms that parameter time courses estimated from incidence data by the AKS method, are indeed able to reassemble the correct dynamics in the ODE solutions. The presented AKS method thus qualifies as potential input estimation approach for ODE models in general.

For comparison, the same ODE simulation was performed by utilizing the alternative $\mathcal{R}_t$ time course calculated from the incidence-based method with fixed serial interval time $s=4$ days and averaging interval $\tau=7$ days, as presented above in Figure~\ref{fig6_germany}C as blue line. Since this methods lacks estimates for parameters $\gamma_t$ and $\theta_t$ as required to solve the ODE system, time course estimates for $\gamma_t$ and $\theta_t$ from the SIRD-AKS method were chosen. The resulting incidence-based ODE trajectories are displayed as blue line in Figure~\ref{fig7_ring}.

In contrast to the ODE trajectories with SIRD-AKS estimates, the ODE solutions with incidence-based $\mathcal{R}_t$ do not agree well with the data. Although, the overall dynamic activity of the data is approximately covered, it does not well capture certain details of the data series, e.g.~the partial drop in the infected flux during the second wave in February 2021. Furthermore, the predicted amplitudes of newly infected individuals are not adequately met, and thus the overall pandemic situation would be underestimated.

It should be noted that the lack of estimates for $\gamma_t$ and $\theta_t$ from the incidence-based method renders some difficulty for a fair method comparison within this ODE-validation approach.

In conclusion, the incidence-based method does not  provide an estimate for the reproduction number $\mathcal{R}_t$ that can be used as an input function in the addressed ODE model. Instead for the shown real-word application, the incidence-based method $\mathcal{R}_t$ generates an ODE solution which differs from the observations in orders of magnitude. In contrast, the SIRD-AKS method reassembles an adequate ODE solution, even in the potentially over-simplified SIRD model.

\section{Conclusion}

The general task of estimating time-dependent parameters in ODE models, also referred to as input function estimation, is known to be conceptually difficult. Either a particular function class, i.e.~a parameterization of a general input function, needs to be provided, or the initial value problem of solving the ODE needs to be transferred into a boundary value problem for which no good optimizers are available \cite{kaschek2012variational}.

To remedy this, we present in this work a non-parametric model-based estimation for all time-dependent parameters in epidemiological ODE models by combining an Expectation-Maximization algorithm with the Augmented Kalman Smoother (AKS) algorithm.

Our approach is able to estimate the time-dependencies in multiple model parameters and independently from the particular input function class.
While other approaches exist that also take advantage of Kalman filter methods for the estimation of a single time-dependency in compartmental models \cite{arroyo2021tracking}, our approach is more general and enables estimation of all time-dependent model parameters simultaneously. 

We showed the performance of our AKS method using an SIRD core model structure (SIRD-AKS) in diverse simulation settings.
The SIRD-AKS performs well even for high noise levels and under different degrees of model misspecification. Thus, the method also yields meaningful estimates in situations where the details of the data generating process remain unknown, meaning that the assumed AKS model structure does not perfectly cover the true process structure.
In particular, the presented approach only utilizes incidence data of infected, recovered and deceased individuals and, as a major advantage compared to existing methods \cite{khailaie2021development,hamouda2020schatzung}, does not require any further prior knowledge on the analyzed disease. 

When applying the SIRD-AKS method to COVID-19 data from Germany, it can be observed that all parameter time course estimates are plausible and in particular the effective reproduction number $\mathcal R_t$ estimate displays the impact of all relevant non-pharmaceutical interventions and applied control measures during the SARS-CoV-2 pandemic in Germany. Further, the simultaneously estimated death rate $\theta_t$ matches well to the time pattern of how many elderly people were infected by the disease.

We further compared our SIRD-AKS method to an alternative approach for the estimation of $\mathcal{R}_t$ which considerably depends on the choice for the value of the assumed Dirac $\delta$-distributed serial interval.

In contrast, the SIRD-AKS method is independent of the distribution of the serial interval and therefore, it does not require any information on the assumed serial interval time as a hyperparameter. Instead, our approach internally copes with the serial interval through the ODE model and is not affected by any inadequate assumptions on the serial interval distribution. Even if the serial interval distribution is also subject to change over time, as for example reported in \cite{ali2020serial}, this is compensated by the general time-dependency e.g.\ of the transmission parameter $\beta_t$ in the SIRD model. 

Compared to the incidence-based method, the SIRD-AKS estimate for the effective reproduction number $\mathcal{R}_t$ shows the pandemic situation more clearly as it is less biased towards $\mathcal{R}_t=1$.
Furthermore, we checked the appropriateness of the estimated parameter time courses by the SIRD-AKS method by plugging them back into the original ODE model. While the incidence-based estimates yielded case numbers that were orders of magnitude below the original data, we were able to correctly reassemble the case numbers by the ODE model from the SIRD-AKS estimates.

In summary, the SIRD-AKS method is a non-parametric model-based estimation approach for the reproduction number and other time-dependent model parameters that only requires incidence data of newly infected, recovered and deceased individuals.
Since by construction the presented approach is not restricted to epidemiological ODE models, it might be used also for input function estimation in non-linear ODE models in general.

\subsection*{Funding}
This work was supported by the German Research Foundation (DFG) under Germany's Excellence Strategy (CIBSS – EXC-2189 – Project ID 390939984) and the German Research Foundation (DFG) through grant 272983813/TRR 179.

\subsection*{Author contributions}
All authors took part in the conception of the study idea. JH developed the method, implemented the algorithm and conducted the analysis under supervision of MR, CT and JT. JH, MR and CT carried out drafting. All authors edited the manuscript and approved the final version for publication.
 
\subsection*{Conflict of interest}
Nothing to declare.

\newpage

\bibliographystyle{apalike}
\bibliography{bibliography}

\begin{thebibliography}{}

\bibitem[Abboubakar et~al., 2021]{abboubakar2021mathematical}
Abboubakar, H., Guidzava{\"\i}, A.~K., Yangla, J., Damakoa, I., and Mouangue,
  R. (2021).
\newblock {Mathematical modeling and projections of a vector-borne disease with
  optimal control strategies: A case study of the Chikungunya in Chad}.
\newblock {\em Chaos, Solitons \& Fractals}, 150:111197.

\bibitem[Ali et~al., 2020]{ali2020serial}
Ali, S.~T., Wang, L., Lau, E.~H., Xu, X.-K., Du, Z., Wu, Y., Leung, G.~M., and
  Cowling, B.~J. (2020).
\newblock {Serial interval of SARS-CoV-2 was shortened over time by
  nonpharmaceutical interventions}.
\newblock {\em Science}, 369(6507):1106--1109.

\bibitem[Arroyo-Marioli et~al., 2021]{arroyo2021tracking}
Arroyo-Marioli, F., Bullano, F., Kucinskas, S., and Rond{\'o}n-Moreno, C.
  (2021).
\newblock Tracking {R of COVID-19}: {A new real-time estimation using the
  Kalman filter}.
\newblock {\em PLoS ONE}, 16(1):e0244474.

\bibitem[Bai et~al., 2020]{bai2020rapid}
Bai, Z., Gong, Y., Tian, X., Cao, Y., Liu, W., and Li, J. (2020).
\newblock {The rapid assessment and early warning models for COVID-19}.
\newblock {\em Virologica Sinica}, 35(3):272--279.

\bibitem[Bani-Yaghoub et~al., 2012]{bani2012reproduction}
Bani-Yaghoub, M., Gautam, R., Shuai, Z., Van Den~Driessche, P., and Ivanek, R.
  (2012).
\newblock Reproduction numbers for infections with free-living pathogens
  growing in the environment.
\newblock {\em Journal of Biological Dynamics}, 6(2):923--940.

\bibitem[Barbarossa et~al., 2020]{barbarossa2020modeling}
Barbarossa, M.~V., Fuhrmann, J., Meinke, J.~H., Krieg, S., Varma, H.~V.,
  Castelletti, N., and Lippert, T. (2020).
\newblock {Modeling the spread of COVID-19 in Germany: Early assessment and
  possible scenarios}.
\newblock {\em PLoS ONE}, 15(9):e0238559.

\bibitem[Brauer et~al., 2012]{brauer2012mathematical}
Brauer, F., Castillo-Chavez, C., and Castillo-Chavez, C. (2012).
\newblock {\em Mathematical Models in Population Biology and Epidemiology},
  volume~2.
\newblock Springer.

\bibitem[Capaldi et~al., 2012]{capaldi2012parameter}
Capaldi, A., Behrend, S., Berman, B., Smith, J., Wright, J., and Lloyd, A.~L.
  (2012).
\newblock Parameter estimation and uncertainty quantication for an epidemic
  model.
\newblock {\em Mathematical Biosciences and Engineering}, 9:553.

\bibitem[Capistr{\'a}n et~al., 2009]{capistran2009parameter}
Capistr{\'a}n, M.~A., Moreles, M.~A., and Lara, B. (2009).
\newblock Parameter estimation of some epidemic models. {The} case of recurrent
  epidemics caused by respiratory syncytial virus.
\newblock {\em Bulletin of Mathematical Biology}, 71(8):1890--1901.

\bibitem[Carrassi and Vannitsem, 2011]{carrassi2011state}
Carrassi, A. and Vannitsem, S. (2011).
\newblock {State and parameter estimation with the extended Kalman filter: An
  alternative formulation of the model error dynamics}.
\newblock {\em Quarterly Journal of the Royal Meteorological Society},
  137(655):435--451.

\bibitem[Chen et~al., 2021]{chen2021numerical}
Chen, X., Li, J., Xiao, C., and Yang, P. (2021).
\newblock {Numerical solution and parameter estimation for uncertain SIR model
  with application to COVID-19}.
\newblock {\em Fuzzy Optimization and Decision Making}, 20(2):189--208.

\bibitem[Contento et~al., 2021]{contento2021integrative}
Contento, L., Castelletti, N., Raimundez, E., Le~Gleut, R., Schaelte, Y.,
  Stapor, P., Hinske, L.~C., Hoelscher, M., Wieser, A., Radon, K., et~al.
  (2021).
\newblock Integrative modelling of reported case numbers and seroprevalence
  reveals time-dependent test efficiency and infection rates.
\newblock {\em medRxiv}.

\bibitem[Cori et~al., 2013]{cori2013new}
Cori, A., Ferguson, N.~M., Fraser, C., and Cauchemez, S. (2013).
\newblock A new framework and software to estimate time-varying reproduction
  numbers during epidemics.
\newblock {\em American Journal of Epidemiology}, 178(9):1505--1512.

\bibitem[Dehning et~al., 2020]{dehning2020inferring}
Dehning, J., Zierenberg, J., Spitzner, F.~P., Wibral, M., Neto, J.~P., Wilczek,
  M., and Priesemann, V. (2020).
\newblock {Inferring change points in the spread of COVID-19 reveals the
  effectiveness of interventions}.
\newblock {\em Science}, 369(6500).

\bibitem[Dong et~al., 2020]{dong2020interactive}
Dong, E., Du, H., and Gardner, L. (2020).
\newblock {An interactive web-based dashboard to track COVID-19 in real time}.
\newblock {\em The Lancet Infectious Diseases}, 20(5):533--534.

\bibitem[Dreano et~al., 2017]{dreano2017estimating}
Dreano, D., Tandeo, P., Pulido, M., Ait-El-Fquih, B., Chonavel, T., and Hoteit,
  I. (2017).
\newblock {Estimating model-error covariances in nonlinear state-space models
  using Kalman smoothing and the Expectation-Maximization algorithm}.
\newblock {\em Quarterly Journal of the Royal Meteorological Society},
  143(705):1877--1885.

\bibitem[Engelhardt et~al., 2016]{engelhardt2016learning}
Engelhardt, B., Fr{\H{o}}hlich, H., and Kschischo, M. (2016).
\newblock Learning (from) the errors of a systems biology model.
\newblock {\em Scientific Reports}, 6(1):1--9.

\bibitem[{Fazit Communication GmbH}, 2021]{website}
{Fazit Communication GmbH} (2021).
\newblock The federal government informs about the corona crisis.
\newblock
  https://www.deutschland.de/en/news/german-federal-government-informs-about-the-corona-crisis,
  Accessed: 2021-10-22.

\bibitem[Ferguson et~al., 2020]{ferguson2020report}
Ferguson, N., Laydon, D., Nedjati~Gilani, G., Imai, N., Ainslie, K., Baguelin,
  M., Bhatia, S., Boonyasiri, A., Cucunuba~Perez, Z., Cuomo-Dannenburg, G.,
  et~al. (2020).
\newblock {Report 9: Impact of non-pharmaceutical interventions (NPIs) to
  reduce COVID19 mortality and healthcare demand}.

\bibitem[Godio et~al., 2020]{godio2020seir}
Godio, A., Pace, F., and Vergnano, A. (2020).
\newblock {SEIR modeling of the Italian epidemic of SARS-CoV-2 using
  computational swarm intelligence}.
\newblock {\em International Journal of Environmental Research and Public
  Health}, 17(10):3535.

\bibitem[Hamouda et~al., 2020]{hamouda2020schatzung}
Hamouda, O. et~al. (2020).
\newblock {Sch{\"a}tzung der aktuellen Entwicklung der SARS-CoV-2-Epidemie in
  Deutschland--Nowcasting}.
\newblock https://edoc.rki.de/handle/176904/6650.4, Accessed 12-11-2021.

\bibitem[Hethcote, 2000]{hethcote2000mathematics}
Hethcote, H.~W. (2000).
\newblock The mathematics of infectious diseases.
\newblock {\em SIAM Review}, 42(4):599--653.

\bibitem[Ho et~al., 2020]{ho2020older}
Ho, F.~K., Petermann-Rocha, F., Gray, S.~R., Jani, B.~D., Katikireddi, S.~V.,
  Niedzwiedz, C.~L., Foster, H., Hastie, C.~E., Mackay, D.~F., Gill, J.~M.,
  et~al. (2020).
\newblock {Is older age associated with COVID-19 mortality in the absence of
  other risk factors? General population cohort study of 470,034 participants}.
\newblock {\em PLoS ONE}, 15(11):e0241824.

\bibitem[Hoertel et~al., 2020]{hoertel2020stochastic}
Hoertel, N., Blachier, M., Blanco, C., Olfson, M., Massetti, M., Rico, M.~S.,
  Limosin, F., and Leleu, H. (2020).
\newblock {A stochastic agent-based model of the SARS-CoV-2 epidemic in
  France}.
\newblock {\em Nature Medicine}, 26(9):1417--1421.

\bibitem[H{\"o}hle and an~der Heiden, 2014]{hohle2014bayesian}
H{\"o}hle, M. and an~der Heiden, M. (2014).
\newblock {Bayesian nowcasting during the STEC O104: H4 outbreak in Germany,
  2011}.
\newblock {\em Biometrics}, 70(4):993--1002.

\bibitem[Hong and Li, 2020]{hong2020estimation}
Hong, H.~G. and Li, Y. (2020).
\newblock {Estimation of time-varying reproduction numbers underlying
  epidemiological processes: A new statistical tool for the COVID-19 pandemic}.
\newblock {\em PLoS ONE}, 15(7):e0236464.

\bibitem[Ivorra et~al., 2020]{ivorra2020mathematical}
Ivorra, B., Ferr{\'a}ndez, M.~R., Vela-P{\'e}rez, M., and Ramos, A.~M. (2020).
\newblock {Mathematical modeling of the spread of the coronavirus disease 2019
  (COVID-19) taking into account the undetected infections. The case of China}.
\newblock {\em Communications in Nonlinear Science and Numerical Simulation},
  88:105303.

\bibitem[Jarvis et~al., 2020]{jarvis2020quantifying}
Jarvis, C.~I., Van~Zandvoort, K., Gimma, A., Prem, K., Klepac, P., Rubin,
  G.~J., and Edmunds, W.~J. (2020).
\newblock {Quantifying the impact of physical distance measures on the
  transmission of COVID-19 in the UK}.
\newblock {\em BMC Medicine}, 18(1):1--10.

\bibitem[Jo et~al., 2020]{jo2020analysis}
Jo, H., Son, H., Hwang, H.~J., and Jung, S.~Y. (2020).
\newblock {Analysis of COVID-19 spread in South Korea using the SIR model with
  time-dependent parameters and deep learning}.
\newblock {\em medRxiv}.

\bibitem[Jones et~al., 2008]{jones2008global}
Jones, K.~E., Patel, N.~G., Levy, M.~A., Storeygard, A., Balk, D., Gittleman,
  J.~L., and Daszak, P. (2008).
\newblock Global trends in emerging infectious diseases.
\newblock {\em Nature}, 451(7181):990--993.

\bibitem[Kalman, 1960]{kalman1960new}
Kalman, R.~E. (1960).
\newblock A new approach to linear filtering and prediction problems.
\newblock {\em Journal of Basic Engineering}, 82(1):35--45.

\bibitem[Kaschek et~al., 2019]{kaschek2019dynamic}
Kaschek, D., Mader, W., Fehling-Kaschek, M., Rosenblatt, M., and Timmer, J.
  (2019).
\newblock {Dynamic modeling, parameter estimation, and uncertainty analysis in
  R}.
\newblock {\em Journal of Statistical Software}, 88(1):1--32.

\bibitem[Kaschek and Timmer, 2012]{kaschek2012variational}
Kaschek, D. and Timmer, J. (2012).
\newblock A variational approach to parameter estimation in ordinary
  differential equations.
\newblock {\em BMC Systems Biology}, 6(1):1--8.

\bibitem[Keeling and Rohani, 2011]{keeling2011modeling}
Keeling, M.~J. and Rohani, P. (2011).
\newblock {\em Modeling infectious diseases in humans and animals}.
\newblock Princeton University Press.

\bibitem[Kermack and McKendrick, 1991]{kermack1991contributions}
Kermack, W.~O. and McKendrick, A.~G. (1991).
\newblock Contributions to the mathematical theory of epidemics--i. 1927.
\newblock {\em Bulletin of Mathematical Biology}, 53(1-2):33--55.

\bibitem[Khailaie et~al., 2021]{khailaie2021development}
Khailaie, S., Mitra, T., Bandyopadhyay, A., Schips, M., Mascheroni, P.,
  Vanella, P., Lange, B., Binder, S.~C., and Meyer-Hermann, M. (2021).
\newblock {Development of the reproduction number from coronavirus SARS-CoV-2
  case data in Germany and implications for political measures}.
\newblock {\em BMC Medicine}, 19(1):1--16.

\bibitem[Kolokolnikov and Iron, 2021]{kolokolnikov2021law}
Kolokolnikov, T. and Iron, D. (2021).
\newblock Law of mass action and saturation in {SIR} model with application to
  coronavirus modelling.
\newblock {\em Infectious Disease Modelling}, 6:91--97.

\bibitem[Kreutz, 2020]{kreutz2020new}
Kreutz, C. (2020).
\newblock A new approximation approach for transient differential equation
  models.
\newblock {\em Frontiers in Physics}, 8:70.

\bibitem[Latsuzbaia et~al., 2020]{latsuzbaia2020evolving}
Latsuzbaia, A., Herold, M., Bertemes, J.-P., and Mossong, J. (2020).
\newblock {Evolving social contact patterns during the COVID-19 crisis in
  Luxembourg}.
\newblock {\em PLoS ONE}, 15(8):e0237128.

\bibitem[Loos et~al., 2018]{loos2018hierarchical}
Loos, C., Krause, S., and Hasenauer, J. (2018).
\newblock Hierarchical optimization for the efficient parametrization of ode
  models.
\newblock {\em Bioinformatics}, 34(24):4266--4273.

\bibitem[Massonis et~al., 2020]{massonis2020structural}
Massonis, G., Banga, J.~R., and Villaverde, A.~F. (2020).
\newblock {Structural identifiability and observability of compartmental models
  of the COVID-19 pandemic}.
\newblock {\em Annual Reviews in Control}.

\bibitem[Merow and Urban, 2020]{merow2020seasonality}
Merow, C. and Urban, M.~C. (2020).
\newblock {Seasonality and uncertainty in global COVID-19 growth rates}.
\newblock {\em Proceedings of the National Academy of Sciences},
  117(44):27456--27464.

\bibitem[Murray, 2002]{murray2002mathematical}
Murray, J.~D. (2002).
\newblock {\em {Mathematical Biology: I. An Introduction. Interdisciplinary
  Applied Mathematics}}.
\newblock Springer.

\bibitem[Nishiura and Chowell, 2009]{nishiura2009effective}
Nishiura, H. and Chowell, G. (2009).
\newblock The effective reproduction number as a prelude to statistical
  estimation of time-dependent epidemic trends.
\newblock In {\em Mathematical and Statistical Estimation Approaches in
  Epidemiology}, pages 103--121. Springer.

\bibitem[Organization and Others, 2014]{world2014vector}
Organization, W.~H. and Others (2014).
\newblock Vector-borne diseases.
\newblock Technical report, WHO Regional Office for South-East Asia.

\bibitem[Prodanov, 2021]{prodanov2021analytical}
Prodanov, D. (2021).
\newblock {Analytical parameter estimation of the SIR epidemic model.
  Applications to the COVID-19 pandemic}.
\newblock {\em Entropy}, 23(1):59.

\bibitem[Rahmandad et~al., 2021]{rahmandad2021behavioral}
Rahmandad, H., Lim, T.~Y., and Sterman, J. (2021).
\newblock {Behavioral dynamics of COVID-19: Estimating under-reporting,
  multiple waves, and adherence fatigue across 92 nations}.
\newblock {\em System Dynamics Review}, 37(1):5--31.

\bibitem[Raim{\'u}ndez et~al., 2021]{raimundez2021covid}
Raim{\'u}ndez, E., Dudkin, E., Vanhoefer, J., Alamoudi, E., Merkt, S.,
  Fuhrmann, L., Bai, F., and Hasenauer, J. (2021).
\newblock {COVID-19} outbreak in wuhan demonstrates the limitations of publicly
  available case numbers for epidemiological modeling.
\newblock {\em Epidemics}, 34:100439.

\bibitem[Rauch et~al., 1965]{rauch1965maximum}
Rauch, H.~E., Tung, F., and Striebel, C.~T. (1965).
\newblock Maximum likelihood estimates of linear dynamic systems.
\newblock {\em AIAA Journal}, 3(8):1445--1450.

\bibitem[Raue et~al., 2013]{raue2013lessons}
Raue, A., Schilling, M., Bachmann, J., Matteson, A., Schelke, M., Kaschek, D.,
  Hug, S., Kreutz, C., Harms, B.~D., Theis, F.~J., et~al. (2013).
\newblock Lessons learned from quantitative dynamical modeling in systems
  biology.
\newblock {\em PLoS ONE}, 8(9):e74335.

\bibitem[Raue et~al., 2015]{raue2015data2dynamics}
Raue, A., Steiert, B., Schelker, M., Kreutz, C., Maiwald, T., Hass, H.,
  Vanlier, J., T{\"o}nsing, C., Adlung, L., Engesser, R., et~al. (2015).
\newblock {Data2Dynamics: {A} modeling environment tailored to parameter
  estimation in dynamical systems}.
\newblock {\em Bioinformatics}, 31(21):3558--3560.

\bibitem[Rockett et~al., 2020]{rockett2020revealing}
Rockett, R.~J., Arnott, A., Lam, C., Sadsad, R., Timms, V., Gray, K.-A., Eden,
  J.-S., Chang, S., Gall, M., Draper, J., et~al. (2020).
\newblock {Revealing COVID-19 transmission in Australia by SARS-CoV-2 genome
  sequencing and agent-based modeling}.
\newblock {\em Nature Medicine}, 26(9):1398--1404.

\bibitem[Schelker et~al., 2012]{schelker2012comprehensive}
Schelker, M., Raue, A., Timmer, J., and Kreutz, C. (2012).
\newblock Comprehensive estimation of input signals and dynamics in biochemical
  reaction networks.
\newblock {\em Bioinformatics}, 28(18):i529--i534.

\bibitem[Soetaert et~al., 2010]{soetart2010deSolve}
Soetaert, K., Petzoldt, T., and Setzer, R.~W. (2010).
\newblock Solving differential equations in {R}: {P}ackage de{S}olve.
\newblock {\em Journal of Statistical Software}, 33(9):1--25.

\bibitem[Staerk et~al., 2021]{staerk2021estimating}
Staerk, C., Wistuba, T., and Mayr, A. (2021).
\newblock {Estimating effective infection fatality rates during the course of
  the COVID-19 pandemic in Germany}.
\newblock {\em BMC Public Health}, 21(1):1--9.

\bibitem[Stapor et~al., 2018]{stapor2018pesto}
Stapor, P., Weindl, D., Ballnus, B., Hug, S., Loos, C., Fiedler, A., Krause,
  S., Hro{\ss}, S., Fr{\"o}hlich, F., and Hasenauer, J. (2018).
\newblock {PESTO: Parameter estimation toolbox}.
\newblock {\em Bioinformatics}, 34(4):705--707.

\bibitem[Sun et~al., 2008]{sun2008extended}
Sun, X., Jin, L., and Xiong, M. (2008).
\newblock Extended {Kalman} filter for estimation of parameters in nonlinear
  state-space models of biochemical networks.
\newblock {\em PLoS ONE}, 3(11):e3758.

\bibitem[T{\"o}nsing et~al., 2018]{tonsing2018profile}
T{\"o}nsing, C., Timmer, J., and Kreutz, C. (2018).
\newblock Profile likelihood-based analyses of infectious disease models.
\newblock {\em Statistical Methods in Medical Research}, 27(7):1979--1998.

\bibitem[Truszkowska et~al., 2021]{truszkowska2021high}
Truszkowska, A., Behring, B., Hasanyan, J., Zino, L., Butail, S., Caroppo, E.,
  Jiang, Z.-P., Rizzo, A., and Porfiri, M. (2021).
\newblock {High-resolution agent-based modeling of COVID-19 spreading in a
  small town}.
\newblock {\em Advanced Theory and Simulations}, 4(3):2000277.

\bibitem[Van~den Driessche, 2017]{van2017reproduction}
Van~den Driessche, P. (2017).
\newblock Reproduction numbers of infectious disease models.
\newblock {\em Infectious Disease Modelling}, 2(3):288--303.

\bibitem[Villaverde et~al., 2019]{villaverde2019full}
Villaverde, A.~F., Tsiantis, N., and Banga, J.~R. (2019).
\newblock Full observability and estimation of unknown inputs, states and
  parameters of nonlinear biological models.
\newblock {\em Journal of The Royal Society Interface}, 16(156):20190043.

\bibitem[Wieland et~al., 2021]{wieland2021structural}
Wieland, F.-G., Hauber, A.~L., Rosenblatt, M., T{\"o}nsing, C., and Timmer, J.
  (2021).
\newblock On structural and practical identifiability.
\newblock {\em Current Opinion in Systems Biology}.

\bibitem[Xin et~al., 2021]{xin2021latent}
Xin, H., Li, Y., Wu, P., Li, Z., Lau, E. H.~Y., Qin, Y., Wang, L., Cowling,
  B.~J., Tsang, T.~K., and Li, Z. (2021).
\newblock {Estimating the Latent Period of Coronavirus Disease 2019
  (COVID-19)}.
\newblock {\em Clinical Infectious Diseases}.
\newblock ciab746.

\bibitem[Yaari et~al., 2013]{yaari2013modelling}
Yaari, R., Katriel, G., Huppert, A., Axelsen, J., and Stone, L. (2013).
\newblock {Modelling seasonal influenza: The role of weather and punctuated
  antigenic drift}.
\newblock {\em Journal of The Royal Society Interface}, 10(84):20130298.

\bibitem[Yanez et~al., 2020]{yanez2020covid}
Yanez, N.~D., Weiss, N.~S., Romand, J.-A., and Treggiari, M.~M. (2020).
\newblock {COVID-19 mortality risk for older men and women}.
\newblock {\em BMC Public Health}, 20(1):1--7.

\end{thebibliography}

\end{document}